\documentclass[prd,twocolumn,showpacs,preprintnumbers,floats,floatfix,apsrev,amsmath,amsfonts,nofootinbib]{revtex4-1}
\usepackage{xcolor}
\usepackage{graphicx}
\usepackage{dcolumn}
\usepackage{txfonts}
\usepackage{multirow}
\usepackage{subcaption}
\usepackage{natbib}
\usepackage{hyperref}
\usepackage{hypernat}
\usepackage{verbatim}
\usepackage{physics}
\usepackage{amsmath}
\DeclareMathOperator\arcsinh{arcsinh}

\newcommand{\lp}{\left(}
\newcommand{\rp}{\right)}
\newcommand{\lc}{\left[}
\newcommand{\rc}{\right]}

\def\ra{\rightarrow}

\def\be{\begin{equation}}
\def\ee{\end{equation}}
\def\ba{\begin{eqnarray}}
\def\ea{\end{eqnarray}}

\def\vf{\varphi}

\def\n{\nu}

\def\ch{{\cal H}}

\def\car{{\mathcal{R}}}

\newcommand{\mpl}{M_{\mathrm{Pl}}}
\newcommand{\LCDM}{\Lambda \mathrm{CDM}}

\newcommand{\kpi}{\frac{k^3}{2 \pi^2}}
\newcommand{\Pe}{\mathcal{P}_{\mathcal{R}}}
\newcommand{\Pt}{\mathcal{P}_{\mathcal{T}}}

\newcommand{\tempo}{\left(t_{s} - t\right)}

\newcommand{\loarg}{\ln\left(\bar{\eta}\right)}
\newcommand{\lomarg}{\ln^{-1}\left(\bar{\eta}\right)}
\newcommand{\losrarg}{\ln^{1/2}\left(\bar{\eta}\right)}
\newcommand{\losarg}{\ln^{2}\left(\bar{\eta}\right)}

\newcommand{\hpsre}{\epsilon_{1}}

\newcommand{\hans}{\mathrm{H}^{(1)}_{\gamma}}
\newcommand{\hant}{\mathrm{H}^{(1)}_{\gamma}}

\begin{document}

\title{A quasi-matter bounce equivalent to Starobinsky inflation}

\author{L.~F.~Guimar\~aes}\email{lfog@cbpf.br}
\affiliation{CBPF - Centro Brasileiro de
	Pesquisas F\'{\i}sicas, Xavier Sigaud st. 150,
	zip 22290-180, Rio de Janeiro, RJ, Brazil.}

\author{F.~T.~Falciano}\email{ftovar@cbpf.br}
\affiliation{CBPF - Centro Brasileiro de
	Pesquisas F\'{\i}sicas, Xavier Sigaud st. 150,
	zip 22290-180, Rio de Janeiro, RJ, Brazil.}

\author{G.~Brando}\email{gbrando@cosmo-ufes.org}
\affiliation{PPGCosmo, CCE - Universidade Federal do Esp\'\i rito Santo, zip 29075-910,
	Vit\'{o}ria, ES, Brazil}

\date{\today}

\begin{abstract}
In this  paper we construct a bounce model that mimics the Starobinsky inflationary model. Our construction relies on Wands' duality, which shows that the Mukhanov-Sasaki equation has a symmetry transformation by changing appropriately its time-dependent mass term. One of the advantages of this constructive method is that one can control every contribution to the primordial power spectrum and check how far we can emulate a given primordial model with a different scenario. In particular, we show that mapping the Starobinsky inflation into a quasi-matter bounce gives the correct relation between the scalar spectral index $n_s-1$ and the tensor-to-scalar ratio $r$.

\end{abstract}



\maketitle

\section{Introduction}

The Planck collaboration~\cite{2018arXiv180706209P, Akrami:2018odb, 2018arXiv180706205P} produced the latest and strongest constraints on the parameters of the standard cosmological model. Their results confirmed previous experiments such as WMAP~\cite{2013ApJS..208...19H, 2013ApJS..208...20B} showing that the 6-parameter $\Lambda$CDM model continues to be the best fit model for the cosmic microwave background (CMB) data at high redshift. In addition, low redshift experiments also point for this concordance model as the best phenomenology to describe the evolution of the universe. Thus, one of the theoretical and observational challenges of present day cosmology is to push our knowledge further back in time into the primordial universe.

Inflation is certainly the most popular model to describe the primordial universe and has become the paradigm dynamics for this era. Inflation alleviates some of the standard model problems such as the flatness problem and provides a consistent origin for the primordial cosmological perturbations~\cite{Guth:1980zm, Linde1982389, Albrecht_inflation, Riotto:2002yw, Lyth:1998xn}. However, there are others competitive models that also solve the same problems of the standard model and have a good fit to the available observational data such as bounce models~\cite{Brandenberger:2012zb, Falciano:2015kya, Falciano:2008gt, Peter:2002cn}, pre-big-bang~\cite{Durrer:2002jn, Enqvist:2001zp}, ekpyrotic~\cite{Khoury:2001zk, Lyth:2001pf, Lehners:2007ac, Martin:2001ue} and string-inspired models~\cite{Brandenberger1989391, String_gas_Battefeld}.

In particular, bounce models figure among the simplest extensions of the standard model. Evidently, in order to produce a bounce one has to violate the null energy condition or to appeal to modified gravity theories. But similarly to inflation, one can also use bounce models as a pure phenomenological scenario. Furthermore, a close analysis shows that the theoretical support for inflation is as good as for bounce models. Therefore, the current status is that there is no reason to privilege one over the other.

Future experiments will allow us to probe deeper into primordial universe physics by measuring the CMB B-mode polarization, primordial non-gaussianities and the spectrum of primordial gravitational waves~\cite{Matsumura:2013aja, Abazajian:2016yjj, Ade:2018sbj, Bouchet:2011ck}. This new data can break the above mentioned degeneracy between different primordial universe scenarios. In this context, an important theoretical challenge is to make concrete predictions that would allow us to to discriminate between these models.

The aim of the present paper is to construct a bounce model that mimics the inflationary model that best fit the observation, namely Starobinsky inflation. Our construction relies on Wands' duality~\cite{Wands:1998yp, Brustein:1998kq, Finelli:2001sr}, which shows that the Mukhanov-Sasaki equation~\cite{Mukhanov:1990me, Brandenberger:2004LNP...646..127B} displays a symmetry transformation by changing appropriately its time-dependent mass term. One of the advantages of this construction is that one can control every contribution to the primordial power spectrum and check how far we can emulate a given primordial model with a different scenario. Thus, the limits of this construction indicate how one can distinguish different primordial universe models. In particular, we show that a quasi-matter bounce can reproduce the same dependence of the scalar spectral index $n_s$ and the tensor-to-scalar ratio $r$ with the slow-roll parameters as happens in Starobinsky inflation but there is a numerical factor that encodes the physical different between these two models.

The paper is organized as follows. Section~\ref{DfCP} briefly review the basic features of linear cosmological perturbation theory both in GR and modified theories of gravity and highlight some important features of Starobinsky's inflation.  In section~\ref{MSI} we introduce Wands' duality and use it to construct the appropriate collapsing phase prior to the bounce. Section~\ref{CB} is devoted to the bounce phase, which is realized by quantum effects within the Loop Quantum Cosmology scenario and in section~\ref{Con} we conclude with some final remarks.

\section{Cosmological Perturbations}\label{DfCP}

In this paper, we are interested in linear cosmological perturbations and how they can be connected with cosmological observations. There are different primordial scenarios and for each of them the cosmological perturbations have specific developments. The original formulation of Starobinsky inflation~\cite{Starobinsky:1980te} is a modified theory of gravity~\cite{DeFelice:2010aj, Hwang:1996bc, Hwang:1996xh} and hence has to be treat differently from the conventional single field inflationary models\footnote{The fact that Jordan frame description of Starobinsky inflation is formally equivalent to a single field inflationary model in GR is a nontrivial exception.} in General Relativity (GR). Notwithstanding, the dynamic equations for the first order perturbations are formally very similar. In this section we briefly summarize the theory of cosmological perturbations for a scalar field in GR and for $f(R)$ theories.

\subsection{Cosmological Perturbations  in General Relativity}
Cosmological perturbation theory shows that at linear order each type of tensor mode evolves independently and hence we can treat scalar and tensor perturbations separately. Let us consider GR minimally coupled with a scalar field in the comoving gauge. Expanding the action up to second order in the curvature perturbation $\car$ \cite{Mukhanov:1990me} gives
\begin{align}
S_{(2)} = \frac{\mpl^2}{2}\int \dd t \, \dd x^3 \, a^3 \frac{\dot{\vf}^2}{H^2} \left[\dot{\car}^2 - \frac{(\partial_{i} \car)^2}{a^2} \right]\ , \label{ac2o}
\end{align}
where a dot means derivative with respect to cosmic time and $ \mpl$ is the reduced Planck mass\footnote{The reduced Planck mass absorbs the $\sqrt{8\pi}$ in its definition, hence $\mpl=m_{pl}/\sqrt{8\pi}=2.44\times10^{18} {\rm GeV}=4.35\times10^{-6} {\rm g}$ and we use throughout the paper $\hbar=c=1$.}. Defining the Mukhanov-Sasaki variable $v(t,\vec{x})$ and the function $z_{s}(t)$ as
\begin{align}
v \equiv z_{s}\car \quad ,\qquad
z_{s} \equiv \frac{a}{H}\sqrt{\rho + p} = a \frac{\dot{\vf}}{H} \ , \label{eqz}
\end{align}
the action \eqref{ac2o} simplifies to
\begin{equation}
S_{(2)} = \frac{\mpl^2}{2} \int \mathrm{d}\eta \, \mathrm{d}x^3 \left[v'^2 -(\partial_{i} v)^2 + \frac{z_{s}''}{z_{s}}v^2 \right]\ ,
\end{equation}
where now the prime means time derivative with respect to conformal time given by $\eta=\int \, a^{-1} \dd t $. Variation of the above action with respect to $v(t,\vec{x})$ gives the Mukhanov-Sasaki equation. Using a Fourier decomposition, the mode function ${v}_{\vb{k}}(\eta)$ satisfies the dynamic equation
\begin{align}
{v}''_{\vb{k}} + \left(k^2 - \mu^2_{s}\right)v_{\vb{k}} = 0\ , \ \mbox{with}\quad \mu^2_{s} = \frac{z_{s}''}{z_{s}}  \label{mesc} \ . 
\end{align}

Eq.~\eqref{mesc} is formally identical to a parametric harmonic oscillator with mass term $\mu_{s}(\eta)$. Its time dependence comes from the background dynamics through the function $z_{s}(\eta)$. Strictly speaking, Wands' duality~\cite{Wands:1998yp} is a variable transformation that leaves this mass term invariant.

The tensor sector of the second order action reads
\begin{equation}
S_{(2)} = \frac{\mpl^2}{8} \int \dd \eta \, \dd x^3 a^2 \left[({h}'_{ij})^2 - (\partial_{l}{h}_{ij})^2\right]\ ,
\end{equation}
where $h_ {ij}(\eta,\vec{x})$ is the tensor part of the metric perturbation, i.~e. a gauge invariant quantity. Using again a Fourier decomposition for each polarization mode $h^{\lambda}_{\vb{k}}(\eta)$ and defining its associated Mukhanov variable 
\begin{equation}\label{mukhtensor}
v^{\lambda}_{\vb{k}} = \frac{a\, \mpl }{2}h^{\lambda}_{\vb{k}}\ ,
\end{equation}
the resulting Mukhanov-Sasaki equation for each polarization is
\begin{align}\label{mstgr}
{v^{\lambda}_{\vb{k}}}'' + \left(k^2 - \mu^2_{t}\right)v^{\lambda}_{\vb{k}} = 0\ , \ \mbox{with}\quad \mu^2_{t} = \frac{a''}{a}   \ . 
\end{align}

It is worth recalling that for a quasi-dust domination where $H^2\approx\dot{\varphi}^2\approx$ constant, both mass term are equal $\mu^2_{s}=\mu^2_{t}$. As a result, the scalar and tensor modes have identical power spectrum ${k}$-dependence.

\subsection{Cosmological Perturbations  in $f(R)$ theories} 
Apart from the degrees of freedom already present in GR, $f(R)$ theories have an extra scalar degree of freedom~\cite{DeFelice:2010aj,Sotiriou:2008rp}. By their formal equivalence with massless scalar-tensor theories, we know that this extra degree of freedom propagates with the speed of light. At the background level, $f(R)$ theories are observationally indistinguishable from the $\LCDM$ model. It is only in the perturbative level that this two frameworks can be put into test.

Considering a FLRW universe, the background value of the Ricci scalar depends only on time. Contrary to GR where the mechanism to generate inflation resides in the matter field (commonly a scalar field with appropriate potential), in $f(R)$ models of inflation it is the non-linearity of the Ricci scalar that guides the evolution without any scalar field. The extra degree of freedom is encoded in $F = \partial f / \partial R$, which can be decomposed as $F(\eta,\vec{x}) = \bar{F}(\eta) + \delta F(\eta,\vec{x})$, where $\bar{F}(\eta)$ is the background and $\delta F(\eta,\vec{x})$ its perturbation. Expanding the action up to second order in $\car$~\cite{DeFelice:2010aj, Hwang:1996bc, Hwang:1996xh} gives
\begin{align}
S_{(2)} &=  \frac{1}{2}\int \dd \eta \, \dd^{3}x \, a^2 Q_{s} \left[{\car'}^2 - (\partial_{i} \car)^2 \right], \label{aperfr} \\
Q_{s} &= 3\mpl^2\frac{{F'}^2/ 2F}{\left[\mathcal{H} + \left(\frac{{F'}}{2F}\right) \right]^2}. \label{defQs}
\end{align}
where $\mathcal{H}\equiv a'/a$ is the Hubble factor in conformal time. The function $Q_{s}$ plays a similar role as $\dot{\varphi}^2/H^2$ in Eq.~\eqref{ac2o}. Therefore, it is straightforward to vary the above action and find 
\begin{align}
&{v}''_{\vb{k}} + \left(k^2 - \mu^2_{fs}\right)v_{\vb{k}} = 0 \quad , \quad v_{\vb{k}} = z_{fs} \, \car_ {\vb{k}}  \quad ,\label{mssfr}\\
& \mu^2_{fs} = \frac{z_{fs}''}{z_{fs}} \quad  , \quad z_{fs} = a \sqrt{Q_{s}}\quad  .\label{defzfs}
\end{align}

Similar to GR, in this scenario the perturbation has quantum origin. Modified theories of gravity follows the same canonical quantization procedure and impose the same Bunch-Davies initial vacuum state for the variable $v$. The scalar power spectrum is defined as 
\begin{align}\label{Pe}
\Pe &= \kpi |\car|^2 = \kpi \frac{\left | v \right |^2}{a^2Q_s} \quad .
\end{align}

The tensor perturbation expansion is completely analogous to GR since there is no extra tensor degree of freedom. Therefore, the definition of the two polarizations remain identical but there is an additional $F$ term multiplying the second order action that now reads
\begin{align}
S_{(2)} = \frac{\mpl^2}{8} \int \dd \eta \, \dd x^3 a^2F \left[({h}'_{ij})^2 - (\partial_{l}{h}_{ij})^2\right]\ .
\end{align}

The extra $F$ term is absorbed in the definition of the mass term. Variation of the action gives
\begin{align}
&{v^{\lambda}_{\vb{k}}}'' + \left(k^2 - \mu^2_{ft}\right)v^{\lambda}_{\vb{k}} = 0 \quad , \quad v^{\lambda}_{\vb{k}} = \frac{\mpl}{2}\, z_{ft} \, h^{\lambda}_{\vb{k}}  \quad ,\label{mstfr}\\
&\mu^2_{ft} = \frac{z_{ft}''}{z_{ft}} \quad ,  \quad z_{ft} = a \sqrt{F}.
\end{align}

Taking into account the polarization states, the spectrum of tensor perturbations is given by
\begin{equation}\label{Pt}
\Pt = 2\times \kpi \left | h \right |^2= \frac{4k^3}{\pi^2\mpl^2} \frac{\left | v \right |^2}{a^2F} \quad .
\end{equation}
The dynamic equation for the scalar and tensor linear perturbation in GR and $f(R)$ theories are formally identical. The difference between Eq.'s~\eqref{mesc},\eqref{mstgr},\eqref{mssfr} and \eqref{mstfr} are encoded in the definition of the Mukhanov-Sasaki variables and their mass terms. During reheating and the bounce phase it is expected that the dynamics to be modified by new phenomena characteristic of these periods. Indeed, loop quantum cosmology corrections modify the formal structure of the dynamic equation depending on an energy scale parameter $\rho_{c}$. We recover the Mukhanov-Sasaki dynamics in the limit $\rho_{c}\rightarrow\infty$.

\subsection{Starobinsky inflation}
Assuming the cold inflationary paradigm, the model that best fit the observation data is the Starobinsky inflation. This model can be described as a single field inflation~\cite{Mukhanov:1990me} (Einstein frame) or as a solution of a modified theory of gravity~\cite{Starobinsky:1980te} (Jordan frame). We shall follow its original formulation and described it in terms of a $f(R)$ gravity using the metric formulation~\footnote{Inflationary models work on both frames~\cite{Mukhanov:2005sc} but physical quantities are well defined only in Jordan frame~\cite{DeFelice:2010aj}. There is an extensive discussion on the validity of the two frames in the literature (see~\cite{Kamenshchik:2016gcy} for more details).}. An exact vacuum de Sitter expansion is a solution of the dynamic equations only if $f(R) = c_0 R^2$~\cite{DeFelice:2010aj}. The Starobinsky inflation propose a theory with $f(R) = R +c_0 {R^2}$, and hence, the gravitational sector of the action reads
\begin{align}\label{acstab}
S = \frac{\mpl^2}{2}\int\dd^4 x \, \sqrt{-g} \left( R + \frac{R^2}{6M^2}\right) \ ,
\end{align}
where $M$ is a mass parameter that gives the energy scale where the dynamics deviates from GR. During the inflationary phase, the scale factor in leading order in the slow-roll parameters can be approximated by~\cite{Starobinsky:1980te,Mukhanov:1990me}
\begin{equation}\label{Stscal1}
a(t) = a_{0}\tempo^{1/2}\exp\left[-\frac{M^2}{12}\tempo^2\right] \ .
\end{equation}

Calculating the next order correction~\cite{Koshelev:2016xqb} gives
\begin{align}
a(t) &= a_{0}\tempo^{-1/6}\exp\left[-\frac{M^2}{12}\tempo^2\right] \ .\label{festab} 
\end{align}

Note that the difference between these two orders is just the power of the polynomial. Since the evolution is dominated by the exponential, this modification is very small. The adequate definition of slow-roll parameters in $f(R)$ theories is slightly different than in GR. Following the nomenclature of~\cite{DeFelice:2010aj} we have
\begin{align}
\hpsre = -\frac{\dot{H}}{H^2} \quad , 
&&
\epsilon_{3} = \frac{\dot{F}}{2HF}\quad ,
&&
\epsilon_{4} =\frac{\ddot{F}}{H\dot{F}}\quad .
\end{align}

Let us calculate these parameters for a scale factor of the form
\begin{equation}\label{starobagener}
a(t) = a_{0}\tempo^{p}\exp\left[-\frac{M^2}{12}\tempo^2\right] \ .
\end{equation}

Straightforward calculation gives
\begin{widetext}
	\begin{align}
	\hpsre &= \frac{6}{M^2\tempo^2}\left(1+\frac{6p}{M^2\tempo^2}\right)\left(1-\frac{6p}{M^2\tempo^2}\right)^{-2}=\frac{6}{M^2\tempo^2}+ \mathcal{O}(M^{-4})\quad ,\label{hpsre}\\ 
	\epsilon_{3} &= -\frac{M^2}{6H^2}\left[1+\frac{p}{H\tempo}\left(1+\frac{3(1-2p)}{M^2\tempo^2}\right)\right]\left[1+\frac{M^2}{6H^2}\left(1-\frac{3p}{M^2\tempo^2}\right)\right]^{-1}
	=-\frac{6}{M^2\tempo^2}+ \mathcal{O}(M^{-4})\quad ,\label{epsilon3}\\
	\epsilon_{4} &=-\frac{M^2}{6H^2}\left[1-\frac{54p(1-2p)}{M^4\tempo^4}\right]\left[1+\frac{p}{H\tempo}\left(1+\frac{3(1-2p)}{M^2\tempo^2}\right)\right]^{-1}
	=-\frac{6}{M^2\tempo^2}+ \mathcal{O}(M^{-4})\quad .\label{epsilon4}
	\end{align}
\end{widetext}

All three slow-roll parameters are equal in leading order and do not depend on the power $p$ of the polynomial in the scale factor \eqref{starobagener}. Any correction from a different $p$ is at least of order $\mathcal{O}(M^{-4})$.

During a quasi-de Sitter expansion, the general solution of Eq.'s~\eqref{mssfr} and \eqref{mstfr} can be written in terms of Hankel functions of order $\gamma$, which depends on the slow-roll parameters. Assuming a Bunch-Davies initial state and following the standard matching procedure at horizon crossing one can show that
\begin{equation}
v_{\vb{k}}(\eta)=\frac{\sqrt{\pi |\eta|}}{2}e^{i(1+2\gamma )\pi/4}H_{\gamma}^{(1)}(k|\eta|)
\end{equation}

The evolution of the scalar perturbations gives~\cite{DeFelice:2010aj}
\begin{align}
\Pe \approx \frac{1}{Q_{s}}\left(\frac{H}{2\pi}\right)^2\left(\frac{|k\eta_c|}{2}\right)^{n_{\car} - 1}&\ ,&\ n_{\car} - 1 \approx -4\hpsre + 2\epsilon_{3} - 2\epsilon_{4}\ ,
\end{align}
where $\eta_c$ is the time when the wave-number $k$ crosses the horizon. The tensor perturbations follows a similar reasoning mutatis mutandis the evolution
\begin{align}
\Pt \approx \frac{2}{\pi^2 F}\left(\frac{H}{\mpl}\right)^2\left(\frac{|k\eta_c|}{2}\right)^{n_{T}}&\ ,&\ n_{T}\approx -2\hpsre - 2\epsilon_{3}\ .
\end{align}
Finally, the tensor-to-scalar ratio reads
\begin{equation}
r\equiv \frac{\Pt}{\Pe}\approx\frac{8 Q_s}{\mpl^2F}\approx 48\epsilon_{3}^2
\end{equation}

It is convenient to express all observables in terms of the number of e-folds. By definition, the total number of e-folds is $N=\log{a_f/a_i}$ where $t_i$ and $t_f$ are respectively, the onset and end of inflation, i.~e.
\begin{align}
N &\approx \frac{M^2}{12}\left(t_{s} - t_i\right)^2\approx\frac{1}{2\hpsre}\approx-\frac{1}{2\epsilon_3}
\end{align}

Combining these results, the spectral index and the tensor-to-scalar ration read
\begin{align} \label{ns_r_star}
n_{\car} - 1 \simeq -4\hpsre = -\frac{2}{N} \quad ,\quad 
r &\simeq 48\epsilon_{3}^2 \simeq \frac{12}{N^2}\quad .
\end{align}

The Planck 2018 release~\cite{Akrami:2018odb} gives a spectral index of $n_\car=0.9649\pm0.0042$ at $68\%$ confidence level. This implies that $50< N< 65$. We can recast the spectral index and the tensor-to-scalar ratio as
\begin{align} 
&n_{\car} - 1 \approx -3,51\times 10^{-2}\left(\frac{N}{57}\right)^{-1} \ ,\label{nsPlanck}\\
&	r \approx 3,69\times 10^{-3}\left(\frac{N}{57}\right)^{-2}\ .\label{rPlanck}
\end{align}

The Planck $95\%$ confident level upper limit on the tensor-to-scalar ratio is $r_{0.002}<0.10$. This value is even tightened by a combining analysis with the BICEP2/Keck Array BK14 data that bring the tensor-to-scalar value to $r_{0.002}<0.064$. The predicted value for the Starobinsky inflation Eq.~\eqref{rPlanck} is safely within the observational measurements.

\section{Mimicking Starobinsky Inflation}\label{MSI}

The Starobinsky model describes a universe with a violent quasi-de Sitter expansion. This primordial universe model has several known advantages that we simply summarize here by stating that it is the inflationary model that best fits the data. It can be considered as the archetype of inflationary models. Thus, in order to be considered as competitive, any primordial universe model must fit the data as well as Starobinsky´s model. 

Our goal now is to construct a bounce model that encodes the key features of the Starobinsky model in the first perturbative order. The suitable mathematical tools for this is Wands' duality~\cite{Wands:1998yp}. This duality can be understood as a symmetry transformation that leaves the mass term of the Mukhanov-Sasaki equation invariant.

All linear order perturbation equation described in section~\ref{DfCP} has the same structure. They are parametric oscillators with time dependent mass terms. The dynamics of the background enters only on the mass $\mu^2_\alpha\equiv z''_\alpha/z_\alpha$ where the index $\alpha$ designates if we are considering a scalar or a tensor perturbation and if the framework is the vacuum $f(R)$ theory or the scalar field minimally coupled in GR. Any two distinct backgrounds composing the same mass term $\mu_\alpha$ will produce the same evolution for the linear order perturbation. In order to implement this idea, consider a given function $z_\alpha(\eta)$. We can define a new function
\begin{equation}\label{wdual}
\tilde{z}_\alpha(\eta)\equiv c_0 \ z_\alpha(\eta)\, \int_{\eta_{*}}^{\eta}\frac{\dd x}{z_\alpha^2(x)} \ ,
\end{equation}
with $c_0$ and $\eta_{*}$ two arbitrary constants. It can be straightforwardly verified that 
\begin{equation}
\tilde{\mu}^2_\alpha(\eta)\equiv \frac{\tilde{z}''_\alpha}{\tilde{z}_\alpha}=\frac{z''_\alpha}{z_\alpha}=\mu^2_\alpha(\eta)\ .
\end{equation}

The arbitrary constant $c_0$ only re-scales the function $z_\alpha$ but has no observational effect, whereas $\eta_{*}$ sets a family of one parameter solutions. Let us consider a specific scenario to exemplify how this duality works. Scalar perturbations with a minimally coupled scalar field in GR are described by Eq.'s~\eqref{eqz} and \eqref{mesc}. An exact de Sitter universe has $\dot{\varphi}/H$ constant, hence, $z_s\propto a\propto -{1}/{\eta}$. Using transformation Eq.~\eqref{wdual} we find that $\tilde{z}_s=\eta^2$, which describes a dust dominated universe. Therefore, an expanding de Sitter universe produce the same mass term for the linear scalar perturbation as a contracting dust dominated universe. As a consequence, both has the same spectrum of solution for the their Mukhanov-Sasaki variable. It is not a coincidence that matter-bounce scenarios produce scale-invariant power spectrum~\cite{Finelli:2001sr,WilsonEwing:2012pu,deHaro:2015wda}. 

Generically, a universe dominated by an adiabatic perfect fluid with equation of state given by $p=\omega\, \rho$ (with constant $\omega$) has a scale factor with a power law in cosmic time of the form $a(t) \propto t^{2/3(1+\omega)}$. In terms of conformal time, the scale factor evolves as $a(\eta) \propto \eta^{\frac{1}{2} - \nu}$ with $\nu = \frac{3}{2} -\frac{3(1+\omega)}{1+3\omega}$. In GR, the function $z_s=a\sqrt{\rho+p}/H\propto a$ and apart from a constant factor it coincides with the $z_t$. Thus, both mass term are given by $\mu^2_s=\mu^2_t=a''/a$. A radiation fluid has zero mass term since $a\propto \eta$ and there is no possible duality to be performed. For all other fluids, the mass term and the power spectrum associated with this evolution are given by
\begin{align}
\mu^2 &= \frac{\nu^2 - 1/4}{\eta^2}\quad , \quad \nu =  \frac{3}{2} -\frac{3(1+\omega)}{1+3\omega}\quad ,\label{eqnuw}\\
\mathcal{P}_{u} &= \frac{C^2(|\nu|)k^2(-k\eta)^{1 - 2|\nu|}}{4\pi^2}\quad ,
\end{align}
where $C^2(|\nu|)$ is a numeric coefficient. Note that the above power spectrum is invariant under $\n \ra - \n$, which can be translated into a transformation of the fluid's equation of state as
\begin{align}\label{dueqsta}
\omega \ra \tilde{\omega} = \frac{1 +\omega}{-1+ 3\omega}\quad .
\end{align}

This transformation has two fixed points at $\omega=-\frac13$ and $1$. For these fixed points, the evolution of the linear perturbations is univocally determined by the background dynamics. For any other value, there are two background dynamics associated with the same perturbed dynamics. Indeed, it is straightforward to verify that two subsequent transformations return to the same equation of state, i.e. $\tilde{\tilde{\omega}}=\omega$. Therefore, in general, there is a pair of adiabatic perfect fluid background dynamics associated with the same evolution for the linear perturbations. Even though de Sitter evolution is not a power law for the scale factor, its duality transformation is still described by Eq.~\eqref{dueqsta}. As already mentioned before, a de Sitter universe, which has $\omega=-1$ is mapped into a dust dominated universe $\omega=0$.

\begin{figure}
	\includegraphics[width=0.45\textwidth,height=5.6cm]{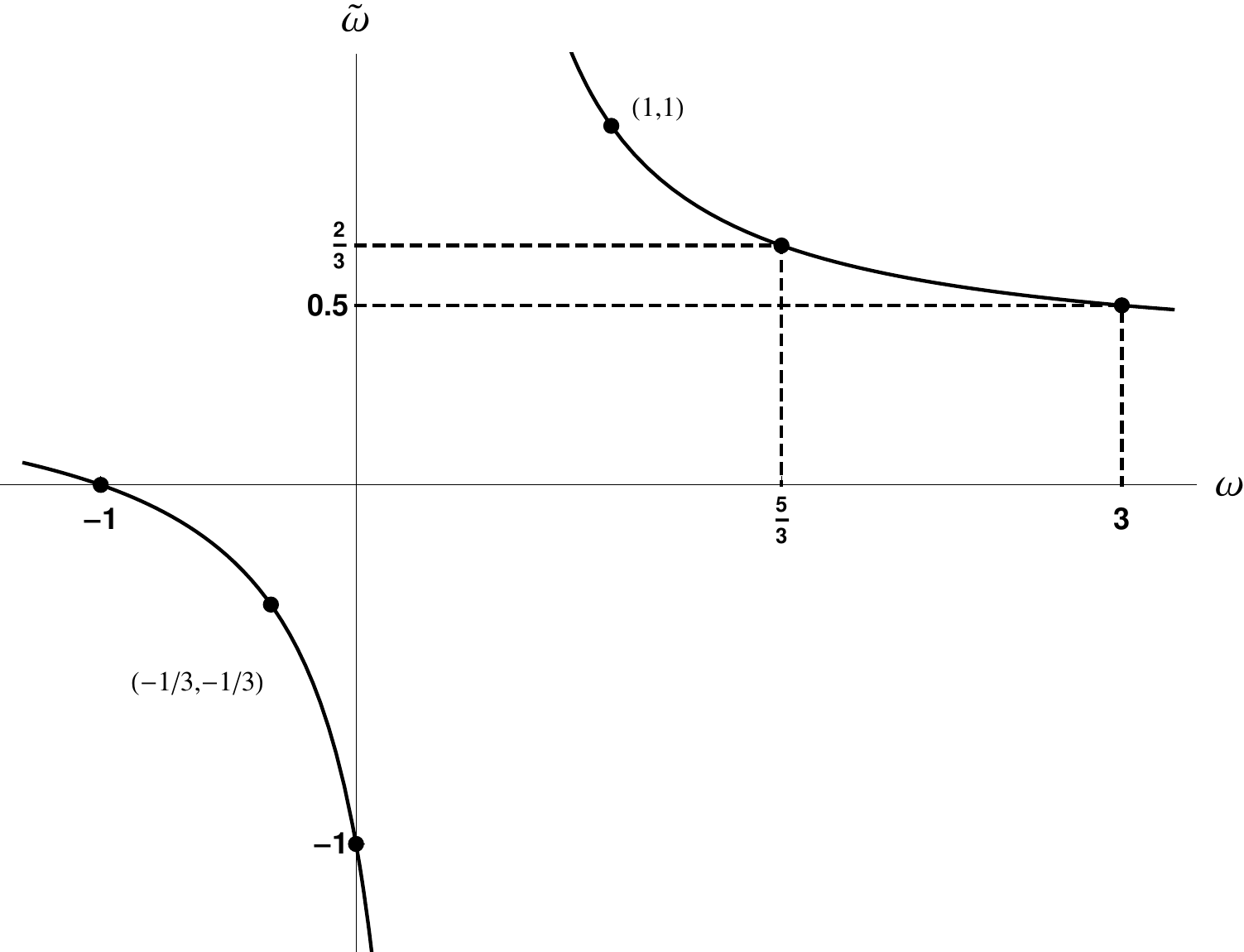}
	\caption{Wands' duality maps an equation of state $\omega$ into $\tilde{\omega}$. There are only two fixed point that mapped into itself given by $\omega=-\frac13$ and $1$. The solid lines represent the map according to Eq.~\eqref{dueqsta}. The dots mark conventional equation of states in cosmology such as $\omega=-1,-\frac13,0,1$.}\label{FigwandsDuality}
\end{figure}

Note, however, that the duality transformation does not specify the theoretical framework. Eq.~\eqref{wdual} map twp distinct $z_\alpha$ functions but does not restrict in which scenario we work with. We can map, for instance, a scalar perturbation in GR into another GR dynamics, $z_{s}\rightarrow \tilde{z}_{s}$, but we can also map a scalar perturbation in GR into a $f(R)$ scenario with the adequate definition of $z_{fs}$, i.e. $z_{s}\rightarrow \tilde{z}_{fs}$.

In the same manner we can map a de Sitter expanding universe into a contracting matter-dominated universe, we shall construct a contracting universe that share the same mass term of the Starobinsky inflation. Though, there is one pitfall. Wands' duality is defined using the conformal time while Starobinsky inflation has an explicitly expression for the scale factor in terms of the cosmic time. The conventional scheme would be to use the definition of conformal time to invert $a(t)$ into $a(\eta)$ but this relation cannot be analytically inverted for Eq.~\eqref{festab}. In order to circumvent this issue, we shall work with a slight modification of Starobinsky's scale factor. As has been argued above, the power of the polynomial in the scale factor is subdominant up to order $\mathcal{O}(M^{-4})$. We shall use this freedom to define a scale factor that allows us to invert the relation and find $a(\eta)$. The appropriate definition of the scale factor which shall be used henceforward is
\begin{align}
a(t) = a_{0} \left(t_{s} - t\right)^{-1} \exp \left[-\frac{M^2}{12}\left(t_{s} - t\right)^2 \right]\quad . \label{fep1}
\end{align}

The associated conformal time is
\begin{align}
\eta &= \int \frac{\mathrm{d}t }{a(t)}=\frac{-6}{a_{0} M^2} \exp \left[\frac{M^2}{12}\left(t_{s} - t\right)^2 \right] \quad .
\end{align}
Thus, the scale factor reads
\begin{equation}
a(\eta) = -\frac{\sqrt{3}}{M\eta}\frac{1}{ \losrarg}\quad .\label{altc}\\
\end{equation}
where we have defined $\bar{\eta} \equiv -{a_{0} M^2\eta}/{6}$, which is a positive quantity. A pure de Sitter universe has $a\propto -1/\eta$, hence, in Starobinsky inflation, the deviation from de Sitter comes from the $\log(\bar{\eta})$ term. Since $M$ is very large, $\log(\bar{\eta})=\log(a_{0} M^2/{6})+\log(-\eta)\approx \log(a_{0} M^2/{6})$, showing that Eq.~\eqref{altc} indeed describes a quasi-de Sitter evolution. Straightforward calculation also gives the Hubble factor and the slow-roll parameters respectively as
\begin{align}
\mathcal{H}(\eta) &= -\frac{1}{\eta}\left(1+\frac{1}{2\loarg}\right)  \quad ,\label{hcaltc}\\
\hpsre=-\epsilon_{3} &=-\epsilon_{4}= \frac{1}{2\loarg}\left[1+\mathcal{O}\left(\frac{1}{\loarg}\right)\right] \quad . \label{e1altc}
\end{align}

Recalling Eq.s~\eqref{defQs} and \eqref{defzfs} we can calculate the $z_s$ function for a scalar perturbation and its associated mass term. Using their definitions we have in leading order
\begin{align}
z_ {s}(\eta) &=-\frac{\sqrt{6}}{\eta} \frac{1}{\loarg}\left[1+\mathcal{O}\left(\frac{1}{\loarg}\right)\right]\quad ,\label{zsF}\\
\mu^2_s&=  \frac{2}{\eta^2}\left[1 + \frac{3}{2}\frac{1}{\loarg} +\mathcal{O}\left(\frac{1}{\losarg}\right)\right]\quad .\label{musF}
\end{align}

Expression Eq.~\eqref{zsF} can be used to construct a contracting $z^B_s$ function that further will be associated with a bounce model. The duality relation Eq.~\eqref{wdual} gives
\begin{align}
z_{s}^{B}(\eta) &= c_0.z_{s}(\eta) \int_{\eta*}^{\eta} \frac{\mathrm{d}{\eta}'}{z_{s}({\eta}')^2}\nonumber \\
&= \frac{c_0}{3\sqrt{6}}\eta^2 \loarg\left[1-\frac2{3\loarg}+\frac{2}{9\losarg}\right]+C\left(\eta_{*}\right)\nonumber \\
&= C_1\eta^2 \loarg\left[1+\mathcal{O}\left(\frac{1}{\loarg}\right)\right]\label{eqconstc}
\end{align}
where $C_1$ is an arbitrary constant. Is is straightforward to check that $z_{s}^{B}$ and $z_{s}$ produce the same mass term $\mu_s$ up to $\mathcal{O}\left({\lomarg}\right)$. Once we have the function $z_{s}^{B}$, we must specify within which scenario the universe is evolving. This extra step is necessary to associate $z_s^{B}$ with a specific background dynamics. For purpose of the present analysis, we choose to immerse this function in a GR contracting solution with the matter content described by a minimally coupled scalar field, hence we have $z_s^{B}=a_B\, \dot{\varphi}/H$.

As argued before, in GR, a quasi-de Sitter inflation is mapped through Wands' duality into a quasi-matter dominated universe. Therefore, we expect that $a^{B}$ should describe a almost matter dominated universe where
\begin{align}\label{approx1}
\dot{\vf}^2 \simeq 2V&& \Rightarrow&& H^2 \simeq \frac{2V}{3 \mpl^2}&& \Rightarrow&& \frac{\dot{\vf}}{\mpl} \simeq \sqrt{3}H\quad .
\end{align}

As a result, the scale factor $a^B$ should be proportional to the function $z_s^B$. Thus, we have
\begin{align}
a_{B}(\eta) &= a_{B0}\,  \eta^2 \ln({\bar{\eta}})\left[1-\frac2{3\loarg}+\mathcal{O}\left(\frac{1}{\losarg}\right)\right] \quad ,\label{fatescesc}\\
\mathcal{H}	&= \frac{2}{\eta}\left[1+\frac{1}{2\ln({\bar{\eta}})}+\mathcal{O}\left(\frac{1}{\losarg}\right)\right] \quad , \label{hctotal}
\end{align}

In order to find the time dependence of the scalar field and its potential, we can use the exact expression
\begin{align}
\vf'^2 &= 2\lp \ch^2 - \ch'\rp\quad ,\label{varphiNoapprox}\\
V &= \frac{\lp 2 \ch^2 + \ch'\rp}{a^2}\quad .\label{VNoapprox}
\end{align}
which is valid for a scalar field with arbitrary potential $V$. The approximation Eq.~\eqref{approx1} is sufficient to argue that $\dot{\varphi}/H$ is constant, while Eq.~\eqref{varphiNoapprox} gives the correct numerical factor for $\varphi'$. Using Eq.s~\eqref{varphiNoapprox} and \eqref{VNoapprox}, the time dependence of the potential and of the scalar field read
\begin{align}
V(\eta) &=\frac{6}{a_{B0}^2}\frac{1}{\eta^6 \log ^2(\bar{\eta})}\left[1+\frac{15}{6\loarg}+\mathcal{O}\left(\frac{1}{\losarg}\right)\right]\quad , \label{Veta}\\
\vf &= -\sqrt{12}\ln \left[\bar{\eta}\ln^{5/12}(\bar{\eta})\right]+\mathcal{O}\left(\frac{1}{\loarg}\right)\quad .\label{phieta}
\end{align}

As a consistency check we can calculate the effective equation of state given by ratio of pressure and energy density, i.e. $\omega\equiv{p}/{\rho}$ Using the above equations we find
\begin{equation}
\omega=\frac{{\varphi'^2}-{2a^2V}}{{\varphi'^2}+{2a^2V}}=-\frac{1}{6\loarg}+\mathcal{O}\left(\frac1\losarg\right)\quad .\label{weff}
\end{equation}

For $\bar{\eta} \gtrsim 10^4$, the equation of state is close to zero with less than $2\%$. Recall that $\bar{\eta}=-a_0M^2\eta/6$ and the mass parameter is expected to be very large, hence relatively small values of conformal time should already satisfy this condition. It is worth noticing that $\omega \lesssim 0$. This is a crucial property to guarantee a slight redshift in the almost scale invariant power spectrum. A positive equation of state would produce a blueshift that contradicts current observation.

\begin{figure}[h]
	\centering
	\includegraphics[width=6.5cm]{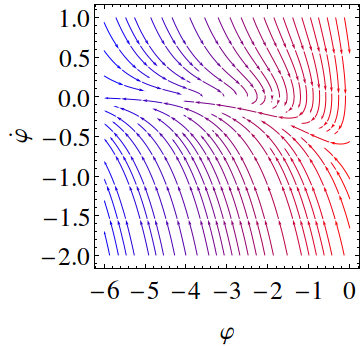}
	\caption{Phase portrait of $\dot{\vf}$ versus $\vf$ for the potential Eq.~\eqref{vphiqmd} using the values $V_0=\varphi_{\ast}=1$. One can see that the dynamics generated by this reconstructed potential is very similar to the exact dust $(p=0)$ potential showing that the square root deformation of the exponential potential work as a small correction.}
	\label{fig:phaseplot}
\end{figure}

Finally, we can combine the above equations to find the potential in terms of the scalar field $V(\varphi)$. After some simple algebra we find
\begin{align}
V(\vf) &=  V_0 \sqrt{1 - \vf/\varphi_{\ast}}\, e^{\sqrt{3} \, \vf}\quad , \label{vphiqmd}
\end{align}

with $V_0$ and $\varphi_{\ast}$ two constant parameters that completely specify the potential. A dust fluid can be described by a scalar field with potential $\exp\left[\sqrt{3} \, \vf\right]$, hence it is not surprising that $V(\vf)$ has this kind of exponential dependence. The novelty is the square root correction, which is intrinsically related to the polynomial correction in the scale factor of Starobinsky inflation. We can again check our construction plotting the phase portrait associated with potential Eq.~\eqref{vphiqmd}. Fig.~\ref{fig:phaseplot} shows the trajectories of the scalar field in the $\left(\varphi,\dot{\varphi}\right)$ plane. For relative large values of $\varphi$ the velocity $\dot{\varphi}$ rapidly goes to zero, which is consistent with a dust fluid given the exponential dependence of the potential $V(\varphi)$.

\section{Crossing the Bounce}\label{CB}

Bounce models are a subclass of nonsingular models that commonly has a single contracting phase followed by an expanding phase. By construction, the contracting phase is smoothly connected to the expanding phase, hence the universe is eternal and free of spacetime singularities. However, this does not mean that one should oppose bounce and inflationary models. Even though a pure inflationary mechanism cannot avoid the initial singularity~\cite{Borde:1994PRL72,Borde_IJMP1996}, a nonsingular model can accommodate an inflationary phase~\cite{Falciano:2007yf, Falciano:2008nk}. However, bounce models are frequently understood as alternative to inflation. 

There are viable bounce models that are consistent with almost scale-invariant power spectrum and small tensor-to-scalar ratio~\cite{NPN_Peter_Pinho_1,Peter:2008qz,Bojowald_RepProgPhys2015,Cai:2014zga,Craig_Singh_CQG2013,Astekar_PRD74,Singh_PRD.74.2006}. In these models, the dynamic through the bounce influences the observable effects. For instance, the mode mixing of scalar perturbations across the bounce is responsible for producing the almost scale-invariant power spectrum. Therefore, it seems reasonable that in order to consider bounce models as a physically viable scenario for the primordial universe, one should recognize them as alternative to inflation and not just as a complementary phase prior to it.

Bounce and inflation have completely distinct background dynamics. Besides the different concerning the singularity problem, at the background level, inflation and bounce models have different shortcomings and theoretical challenges of their own~\cite{Brandenberger:2000as,Jerome_PhysRevD.63.123501,Battefeld:2014uga,Brandenberger:2009jq}. Notwithstanding, at first order perturbation, bounce and inflation are formally very similar. Indeed, Wands' duality described in section~\ref{MSI} is one manifestation of the mathematical similarity between these two scenarios.

Generically, the dynamics of linear perturbations $\nu_{\vb{k}}$ are described by a parametric oscillator equation like Eq.~\eqref{mesc} where the time-dependent mass term $\mu_\alpha$ encodes the background dynamics. In each case we have a specific definition for $\nu_{\vb{k}}$ and $\mu_\alpha$ but the framework is almost identical. Let us compare some of their features.

In both scenarios, even though for different physical reasons, the initial conditions are set in the most (possible) remote past and have a quantum vacuum fluctuation origin. In inflationary models, we have a quasi-de Sitter expansion, which makes the physical length of interest for present cosmology much smaller than the curvature scale. As a consequence, the perturbations are not influenced by the expansion and the initial state is set as a Minkowski vacuum state. In a bounce model, the initial conditions are given in the far past much before the bounce phase. The universe is immense and with negligible curvature, hence, the initial state is a Bunch-Davies vacuum.

As the universe evolves the relation between the physical length and the Hubble length changes. In both scenarios the ratio between these two lengths increases. In terms of the perturbed dynamic equation, this means that with the background evolution, the mass term increases compared to the wavenumber until they become comparable in magnitude. This moment specifies the crossing from outside to inside the potential for the perturbations. The mass term continues to grow until it reaches a maximum that typically locates the bounce or the reheating period for inflationary models. Then the potential starts to decrease until its value becomes again comparable to the wave number characterizing the crossing outside the potential (inside the Hubble length)\footnote{Note that the description in term of the potential for the perturbation (the time dependent mass term) is the opposite as compared to the relative size of the physical and Hubble lengths. Crossing outside the Hubble length means going inside the potential and vice-verse.}. Thereupon, both scenarios are connected to the FLRW radiation epoch and the dynamics follows the standard model.

It is evident from the above description that the violent quasi-de Sitter expansion phase is related to the long contracting phase of bounce models. Moreover, the reheating phase of inflation should be compare to the physical processes during the bounce phase. Thus, it is not surprising that the reheating and the bounce are the two most speculative periods of the evolution.

Inflationary models often overlook the details of the reheating processes. In a certain sense, this is due to the assumption that whatever physical process taking place in this period should only transfer energy into the matter fields and not significantly modify the other physical quantities such as the almost scale-invariant power spectrum or the tensor-to-scalar ratio\footnote{It is worth mentioning that non-gaussianities encoded in the bispectrum are much more sensitive to reheating.}. This idea has support on Weinberg's theorem~\cite{Weinberg:2003sw} that states that, in the large wavelength limit, the field equations for the cosmological perturbations in the Newtonian gauge always have an adiabatic solution with $\car$ constant and nonzero in all eras. 

In contrast, bounce models can not avoid examining the bounce phase since one must define the physical mechanism that produces the bounce. In addition, the physics of the bounce remains encoded in the spectrum of primordial perturbations. As we will show in the following, the relation between the scalar spectral index and the tensor-to-scalar ratio depends on the physics of the bounce. The observational data available are not yet sensitive enough to discriminate between different bounce mechanisms but as in the case of non-gaussianities, future experiments might allow us to probe the physics of the bounce.

In order to connect the contracting phase of the model constructed in the last section to the CMB observables, in the following sections we shall describe the bounce as a quantum gravity effect using the Loop Quantum Cosmology (LQC) framework~\cite{WilsonEwing:2012pu,WilsonEwingLCDM}. There are other appealing frameworks such as Wheeler-DeWitt~\cite{Pinto-Neto:2013toa, PintoNeto:2012ug, Vitenti:2012cx, Falciano_PRD_2015, PintoNeto:2005gx, Falciano:2013uaa} or string cosmology~\cite{Brandenberger_2011, Gasperini:2002bn}. However, LQC has analytical bounce solutions for a scalar field mimicking a perfect fluid, hence, from a technical point of view, it is the most direct description to accommodate a previous Starobinsky-like contracting phase.

Loop quantum gravity (LQG) is a non-perturbative, background independent quantum theory of gravity. It is based on a reformulation of GR in terms of the Ashtekar-Barbero variables. The classical variables promoted to operators are the holonomies of the Ashtekar connection and the fluxes of the densitized triads. One important kinematical result of this quantization procedure is the discretization of spacetime, which in turn establishes a minimum of length, area and volume. LQC relies on using loop quantization techniques to quantize the holonomies and the fluxes of homogeneous and isotropic universes. It is not a full quantum gravity theory but an effective approach that hopefully captures the essential features of LQG in a cosmological scenario (for further details see~\cite{AshtekarSingh2011,BojowaldPRD93,Bojowald_RepProgPhys2015,AshtekarPRD73}). The cosmological dynamics can be described by a phenomenological Hamiltonian. Given a flat FLRW metric, the dynamics with respect to cosmic time reads
\begin{align}
H^2&=\frac{\mpl^{-2}}{3}\rho\left(1-\frac{\rho}{\rho_{c}}\right)\quad ,\label{LQCF1}\\
\dot{H}&=-\frac{\mpl^{-2}}{6}(\rho+p)\left(1-\frac{2\rho}{\rho_{c}}\right)\quad ,\label{LQCF2}\\
\dot{\rho}&+3H\left(\rho+p\right)=0\label{LQCEC}\quad ,
\end{align}
where $\rho_{c}$ is a critical energy density that establishes the energy scale where quantum corrections are important\footnote{We have used the conservation of energy-momentum as our third dynamic equation but we could instead have used the Klein-Gordon equation for the scalar field. The two system of equations are equivalent.}. This dynamic system has analytical bounce solutions for perfect fluids $p=\omega\rho$ with constant $\omega$~\cite{WilsonEwing:2012pu,MielczarekPLB3,WilsonEwing2}. Furthermore, we can use a scalar field with exponential potentials to model the perfect fluid. Indeed, using the fact that
\[
\rho=\frac12\dot{\varphi}^2+V(\varphi)\quad , \quad p=\frac12\dot{\varphi}^2-V(\varphi)\quad ,
\]
one can show that there is an exact solution 
\begin{align}
\rho&=\rho_c \left(\frac{a_B}{a}\right)^{3(1+\omega)}\quad ,\label{LQCsol1}\\
a(t)&=a_B\left(1+
\alpha^2\left(t-t_B\right)^2\right)^{1/3(1+\omega)}\label{LQCsol2}\quad ,\\
\varphi(t)-\varphi_B&=\frac{\sqrt{\rho_c(1+\omega)}}{\alpha}
\arcsinh\bigg(\alpha(t-t_B)\bigg)\quad , \label{LQCsol3}
\end{align}
where $\alpha = \sqrt{3\rho_{c}} (1 + \omega)/2 \mpl$. The parameters $t_B$ and $a_B$ are respectively the values of the cosmic time and the scale factor at the bounce. Note that the energy density  reaches its maximum value at the bounce Eq.~\eqref{LQCsol1}. This is a characteristic feature of symmetric bounces. The scalar field potential for this solution is given by
\begin{equation}\label{VLQC}
V=\frac{\rho_c (1-\omega)}{2}\sech^2\left[
\frac{\alpha\left(\varphi-\varphi_B\right)}{\sqrt{\rho_c(1+\omega)}}
\right]
\quad ,
\end{equation}
where $\varphi_B$ is an arbitrary constant. This solution has two parameters $a_B$ and $\varphi_B$ in addition to the energy density scale $\rho_c$ of LQC. The classical limit is approached in the $\rho_c\rightarrow\infty$. In this limit, Eq.~\eqref{VLQC} tends to $V\sim \exp\left(\sqrt{3(1+\omega)}\varphi/\mpl\right)$, which corresponds to the scalar field potential that describes a perfect fluid with equation of state $\omega$ in GR.

\subsection{Scalar Perturbations in Bounce Models}\label{ScaPert}

Quantum cosmology is an attempt to include quantum effects in the evolution of the universe. In this manner, we must necessarily consider modifications in the GR equations of motion. However, bounce models generically assume that far from the bounce region we recover the GR dynamics. Therefore, long before and after the bounce the scalar perturbations are described by
\begin{align}
{v}'' + \left(k^2 - \mu^2_{s}\right)v = 0\ , \ \mbox{with}\quad \mu^2_{s} = \frac{z_{s}''}{z_{s}}  \label{mesc2} \ ,\\
v \equiv z_{s}\car \quad ,\qquad z_{s} \equiv \frac{a}{H}\sqrt{\rho + p} = a \frac{\dot{\vf}}{H} \ . \label{eqz2}
\end{align}

Using the quasi-matter dynamics of last section Eq.~\eqref{eqconstc}, we find that the classical contracting phase has
\begin{align}
v^{in}(\eta) &= \sqrt{\frac{-\pi \eta}{4}}\, \hans(-k\eta) \label{sclass} \\
\gamma &= \frac{3}{2} + \epsilon_c = \frac{3}{2} + \frac{1}{\loarg} + \frac{2}{3}\frac{1}{\loarg^2}\ ,
\end{align}
where $\hans$ is the Hankel function of the first kind and we have defined in the last expression $\epsilon_c\equiv\gamma - \frac{3}{2}$. The $\epsilon_c$ will play a role analogous to a slow-roll parameter, which differs from the matter bounce parameter\cite{Elizalde:2014uba} by being a small quantity $|\epsilon_c|\ll 1$. Indeed, during the period of validity of the above solution, this term is very small compared to unit, hence, we can consider series expansion in its powers. Our task now is to describe the bounce and use matching conditions to connect this contracting phase with the expanding phase of the standard model. The LQC perturbed equations have two modifications with respect to GR. The Mukhanov-Sasaki equation now reads
\begin{align}
v'' + \left[\lp 1 - \frac{2\rho}{\rho_{c}} \rp k^2  - \frac{z''}{z}\right]v = 0 \quad ,\label{ms_lqc}
\end{align}
where the $z$ function is defined as
\begin{align}
z = \frac{a \sqrt{\rho + P}}{ H} = \mpl\sqrt{\frac{3(1 + \omega)}{1 - \rho/\rho_{c}}}\  a
\quad . \label{zlqc}
\end{align}

Far away from the bounce, the energy density is much less than the critical density, i.e. $\rho/\rho_c\ll 1$ and we recover the classical definitions. Thus, during the contracting phase far away from the bounce, we have~\eqref{sclass}. We need to match this solution with a solution valid during the bounce. Eq.~\eqref{ms_lqc} can be transformed into an integral equation given by
\begin{align}
v(\eta) = &B_{1}z + B_{2}z \int^{\eta} \frac{\dd \bar{\eta}}{z^2} - k^2 \int^{\eta} \frac{\dd \bar{\eta}}{z^2} \int^{\bar{\eta}} \dd \bar{\bar{\eta}} \, z \,v  \nonumber \\
&+ \frac{2 k^2}{\rho_{c}} \, z \int^{\eta} \frac{\dd \bar{\eta}}{z^2} \, \int^{\bar{\eta}} \dd \bar{\bar{\eta}} \, z \, v\quad .\label{key}
\end{align}

Close to the bounce, it is  the mass term that dominates hence we can series expand the solution in powers of the wave number. The solution Eq.s~\eqref{LQCsol2} and \eqref{LQCsol3} are given in cosmic time. We can interpret the conformal time of the above expression as a function of cosmic time. Using the LQC background solution we find at leading order
\begin{align}
v(t) =&  
B_{1}z(t)+
B_{2}z(t)\lp \frac{a_B^{-3}\mpl^{-2}}{3(1+\omega)}\rp\times \label{slqcq}
\\& \quad \times 
\left[\frac{\alpha^2t^3}{3}{}_2F_1\left[\frac32,\frac{2+\omega}{1+\omega},\frac52,-\alpha^2t^2\right]+c_2
\right]
\ ,\quad \nonumber
\end{align}
where ${}_2F_1\left[a,b,c,z\right]$ is the hypergeometric function and $c_2$ is an integration constant that can be chosen conveniently to simplify the matching at the contracting phase. The function $x^3{}_2F_1\left[\frac32,\frac{2+\omega}{1+\omega},\frac52,-x^2\right]$ goes to a constant in the limit $x\rightarrow\pm\infty$, hence we can choose $c_2$ to cancel this constant term in the far past. Consequently, we will have $2c_2$ in the far future after the bounce. Taking the limit $\alpha t\rightarrow -\infty$ we find that $c_2=\frac{\pi}{4\alpha}$. The coefficient $B_{1}$ represents the decreasing mode during Hubble crossing in the contracting phase. We can immediately see from the above expression that due to the behavior of the hypergeometric function the bounce produces a mode mixing transferring the coefficient $B_2$ to the dominant mode after the bounce.

The validity of the contracting solution Eq.~\eqref{sclass} relies on $\epsilon_c$ being almost constant in time and small $\epsilon_c\ll 1$. Thus, we can perform the matching between the contracting phase and the bounce solution well inside the potential for the perturbation but still very far from the bounce. This means that we should take the limit $k\eta\rightarrow0$ in Eq.~\eqref{sclass} and the limit $t\ll -1/\alpha$ in Eq.~\eqref{slqcq}. In addition, our contracting phase has equation of state given by Eq.~\eqref{weff}, hence we must identify $\omega=-\frac16\epsilon_c$. In this limit we can write the scale factor and the cosmic time in terms of the conformal time, i.e.
\begin{align}
a(\eta) &= a_B\lc \frac{\alpha(1-\epsilon_c/3)}{3} a_B\eta\rc^{2+\epsilon_c}\quad ,\\
\alpha t(\eta) &= \lc \alpha \lp \frac{1 -\epsilon_c/3}{3}\rp a_B\eta \rc^{3 +\epsilon_c} \ ,\label{tc_lqc}
\end{align}
where we have used $\epsilon_c\ll 1$ and kept only the leading order terms. Using Eq.'s~\eqref{zlqc}-\eqref{slqcq} we find
\begin{align}
z(\eta) =&a_B^{3+\epsilon_c}\lp 1 - \frac{13\epsilon_c}{12}\rp\frac{\sqrt{\rho_c}}{2}  \left(\frac{\rho_{c}}{12\mpl^2}\right)^{(1+\epsilon_c)/2} \eta^{2+\epsilon_c} \quad  ,\\
v(\eta) =& B_{1}a_B^{3+\epsilon_c}\lp 1 - \frac{13\epsilon_c}{12}\rp\frac{\sqrt{\rho_c}}{2}  \left(\frac{\rho_{c}}{12\mpl^2}\right)^{(1+\epsilon_c)/2} \eta^{2+\epsilon_c}
\nonumber \\&
\quad -\frac{4B_{2}}{\sqrt{3}}\frac{a_B^{-3-\epsilon_c}}{\sqrt{\rho_c}}
\lp 1 + \frac{5\epsilon_c}{12} \rp
\left(\frac{\mpl^2}{\rho_c}\right)^{(1+\epsilon_c)/2}
\eta^{-1-\epsilon_c}\quad .  \label{vqc_lqc}
\end{align}

This solution has to be matched with the contracting solution Eq.~\eqref{sclass} in the limit $k\eta \ll 1$, namely
\begin{align}
v^{in}(\eta) = \frac{1}{3\sqrt{2}}k^{3/2+\epsilon_c}\eta^{2+\epsilon_c}+\frac{i}{\sqrt{2}}k^{-3/2-\epsilon_c}\eta^{-1-\epsilon_c}\ .
\end{align}
Straightforward comparison shows that
\begin{align}
B_{1} &= 
\frac{\sqrt{2}a_B^{-3-\epsilon_c}}{3\sqrt{\rho_c}}\lp 1 + \frac{13\epsilon_c}{12}\rp \left(\frac{\rho_{c}}{12\mpl^2}\right)^{-(1+\epsilon_c)/2}k^{3/2+\epsilon_c}\ ,
\label{b1lqc}\\
B_{2} &=
-i\frac{\sqrt{3}}{4\sqrt{2}}\frac{\sqrt{\rho_c}}{a_B^{-3-\epsilon_c}}\lp 1 - \frac{5\epsilon_c}{12} \rp\left(\frac{\rho_c}{\mpl^2}\right)^{(1+\epsilon_c)/2}k^{-3/2-\epsilon_c}
\label{b2lqc}\ .
\end{align}

The solution Eq.\eqref{slqcq} is valid across the bounce. Having defined the coefficients $B_1$ and $B_2$ we can find the solution after the bounce. The expanding phase solution is described by taking the limit $t\gg 1/\alpha$ in Eq.\eqref{slqcq}, i.e.
\begin{align}
v^{out}(\eta)&= 
\lc B_{1}
+B_{2}
\left(
\frac{\pi a_B^{-3}(1+\epsilon_c/3)}{3\sqrt{3\rho_c} \mpl }
\right)\rc z(\eta)\\
&=\lc \frac{k^{3/2+\epsilon_c}}{3\sqrt{2}}
-i\frac{\pi \lp 1 - \frac{7\epsilon_c}{6}\rp}{48\sqrt{6}}
\left(\frac{a_B^2\rho_c}{\mpl^2}\right)^{3/2+\epsilon_c}
k^{-3/2-\epsilon_c}\rc \eta^{2+\epsilon_c}
\quad .\nonumber
\end{align}

In cosmological perturbations we are interested in the small wavenumber limit, hence for very small wavenumber it is the $k^{-3/2}$ that dominates. However, this is true only if the numerical factors are of order one. The parameter $\rho_c$ is expected to be smaller but comparable in at least a few order of magnitude of the Planck energy density, i.e. $\rho_c=10^{-n}\rho_{_{\rm Pl}}$, with $1<n<10$. The value of the scale factor at the bounce must be at least a few order of magnitude higher than the Planck mass, otherwise we could not rely on our quantum cosmology effective scenario, i.e. $a_B=10^m \, l_{_{\rm Pl}}$ with $5>m>2$. The ratio between the two term above is 
\begin{align}\label{conds}
&\approx 14.28\times 10^{3(m-n/2)} \, l_{_{\rm Pl}}^{-3} k^{-3}\gg 1\quad .
\end{align}
Therefore, it is indeed the $k^{-3/2}$ the dominant coefficient for all values of interest of wavenumber in cosmology and the scalar perturbation is
\begin{align}
\car&=\frac{v}{z}\approx \frac{\pi}{12\sqrt{2}}\sqrt{\frac{\rho_c}{\mpl^4}}k^{-\frac32-\epsilon_c}\approx 0.185
\sqrt{\frac{\rho_c}{\mpl^4}}k^{-\frac32-\epsilon_c}\quad , \label{nuout}
\end{align}
with spectral index given by
\begin{align}
n_{s} - 1 = -2\epsilon_c\quad . \label{spec_ind_lqc}
\end{align}

As expected, the power spectrum is almost scale invariant but with a small redshift. Using the Planck 2018 release $n_s=0.9649\pm0.0042$ (see Ref.~\cite{Akrami:2018odb}), we have $0.0196<\epsilon_c<0.0155$.

\subsection{Tensor Perturbations}\label{TensPert} 

Similarly to the scalar perturbations, the dynamic equation for tensor perturbations in LQC has quantum corrections proportional to $\rho/\rho_c$. The Mukhanov-Sasaki variable is defined in terms of the tensor perturbations $h=2 v /  z_{T} \mpl$, where function $z_T$ is also modified due to quantum corrections. The Mukhanov-Sasaki equation reads
\begin{align}
v'' + \left[\lp 1 - \frac{2\rho}{\rho_{c}} \rp k^2  - \frac{z_{T}''}{z_{T}}\right] v = 0 \quad ,\label{tms_lqc}
\end{align}
where the function $z_T$ is given by
\begin{align}
z_{T} = \frac{a}{\sqrt{1 - 2\rho/\rho_{c}}}\quad . \label{zt_lqc}
\end{align} 
The tensor perturbations in the contracting phase has the same solution as the scalar perturbations, namely
\begin{align}
v^{in}(\eta) = \sqrt{\frac{-\pi \eta}{4}}\, \hant(-k\eta)
\end{align}
where again $\gamma=3/2+\epsilon_c$. Following the same procedure as before, we can transform the differential equation into an integral equation for $\mu$ similar to Eq.~\eqref{key}. The solution across the bounce can be obtained by a series expansion on powers of the wavenumber. At leading order in $k$, the formal solution to its integral form is
\begin{align}\label{mu_lqc_formal}
v(t) = D_1 \, z_{T}(t) + D_2 \, z_{T}(t) \int^{\bar{\eta}} \frac{\dd \eta}{z_{T}(\eta)^2}
\end{align}
where $D_1$ and $D_2$ are two constants of integration. By virtue of Eq.~\eqref{zt_lqc}, the formal solution is 
\begin{align}
v(t) &= D_1 \, z_{T}(t) + \frac{D_2}{a_B^3} \, z_{T}(t) \lc \frac{\alpha^2 t^3}{3} \, _{2}F_{1} \lp \frac{3}{2}, \frac{2+\omega}{1+\omega}, \frac{5}{2}, -\alpha^2 t^2 \rp  \right. + \nonumber \\
&\left.- t \times\ _{2}F_{1} \lp \frac{1}{2}, \frac{2+\omega}{1+\omega}, \frac{3}{2}, -\alpha^2 t^2\rp + C \rc \quad .\label{fst_lqc}
\end{align}

As before, we chose the constant $C$ conveniently to cancel the constant term in the far past. As a result we have $C=-\frac{\pi\omega}{2\alpha}$. Recall that $\alpha = \sqrt{3\rho_{c}} (1 + \omega)/2 \mpl$ and $\omega=-\frac16\epsilon_c$. In order to match this solution with the contracting phase, we must take the limit $t\ll -1/\alpha$ that gives 
\begin{align}
v(t) =&\frac{D_1 \lp 1 - \epsilon_c\rp}{a_B^{-3-\epsilon_c}} \lp \frac{ \rho_{c}}{12\mpl^2} \rp^{1+\epsilon_c/2} \eta^{2 +\epsilon_c} + \nonumber \\
&\qquad + \frac{D_2}{3a_B^3} \lp 1+\frac{2\epsilon_c}{3} \rp \lp \frac{12\mpl^2}{ \rho_{c}} \rp^{1 +\epsilon_c/2}  \eta^{-1-\epsilon_c}\quad .
\label{muqc_lqc}
\end{align}

This expression has to be matched with the limit $k\eta \ll 1$ for the classical solution~\eqref{sclass}, namely
\begin{align}
v^{in}(\eta) =&  \frac{1}{3\sqrt{2}}k^{\frac32+\epsilon_c} \eta^{2+\epsilon_c}
- i \frac{1}{\sqrt{2}} \,   k^{-\frac32-\epsilon_c} \eta^{-1-\epsilon_c}\ .
\end{align}
Thus, we identify
\begin{align}
D_1 &=  \frac{\lp 1 + \epsilon_c\rp}{3\sqrt{2}a_B^{3+\epsilon_c}}\lp \frac{12\mpl^2}{ \rho_{c}} \rp^{1+\epsilon_c/2}k^{\frac32+\epsilon_c}\quad ,\label{D1}\\
D_2 &=-i\frac{3a_B^3}{\sqrt{2}}
\lp 1-\frac{2\epsilon_c}{3} \rp \lp \frac{\rho_{c}}{12\mpl^2} \rp^{1 +\epsilon_c/2} k^{-\frac32-\epsilon_c}\ .\label{D2}
\end{align}

The expanding phase is given by taking the limit $t\gg1/\alpha$. Thus, we have 
\begin{align}
v^{out} (\eta)&= \lc D_1 - \frac{D_2}{a_B^3}\frac{\pi\mpl}{3\sqrt{3\rho_{c}}}\epsilon_c   \rc \, z_{T}^{c}(\eta) \label{mu_grow}
\end{align}
where $D_1$ and  $D_2$ are given by Eq.s~\eqref{D1} and  \eqref{D2}. It is worth noting that the term proportional to $D_2$ is linear in $\epsilon_c$, hence the mode mixing in the tensor perturbation depends on how small is the slow-row parameter. To leading order in wavenumber, the tensor perturbation reads
\begin{align}
h&=\frac{2v}{z_T\mpl} = \frac{2}{\mpl}\lc D_1 - \frac{D_2}{a_B^3} \frac{\pi\mpl}{3\sqrt{3\rho_{c}}}\epsilon_c  \rc \\
&\approx \frac{i\pi}{6\sqrt{6}} \epsilon_c \sqrt{ \frac{\rho_{c}}{\mpl^4}} k^{-\frac32-\epsilon_c}
\approx 0.214 i \epsilon_c \sqrt{ \frac{\rho_{c}}{\mpl^4}} k^{-\frac32-\epsilon_c}
\quad .\label{muout}
\end{align}

Thus, the tensor spectral index is $n_t=-2\epsilon_c=n_s-1$. Finally, using Eq.s~\eqref{nuout} and \eqref{muout}, we find the tensor-to-scalar ratio
\begin{align}
r&=\frac{\Pt}{\Pe}=2\frac{|h|^2}{|\car|^2}=\frac83 \epsilon_c^2=\frac23 (n_s-1)^2 \label{r_lqc}
\end{align}

Note that we succeed in obtaining the same relation between $n_s-1$ and $r$ as in the Starobinsky inflation. However, even though with the correct power of the slow-roll parameter $\epsilon_c^2$, there is a numerical factor difference of order unit. Eq.~\eqref{ns_r_star} shows that Starobinsky inflation has a relation between the scalar spectral index and the tensor-to-scalar ratio given by
\begin{equation} \label{rstab_conc}
r = 3\lp n_s - 1 \rp^2 \quad ,
\end{equation}
hence our model is a factor $2/9$ smaller. This difference is a convolution of two contributions coming from the ratio $z_{s}/z_{T}$ but they have completely distinct physical origin.

First, in inflationary models, the ratio $\left(z_{s}/z_{T}\right)^2$ is $1/2\epsilon_{c}^{2}$ larger than its value in bouncing models. Indeed, one can check that in the Starobinsky model we have $\left(z_{s}/z_{T}\right)^2=Q_s/F\approx \frac32 \mpl^2\epsilon_{c}^2$, while for a matter bounce model we have $\left(z_{s}/z_{T}\right)^2=3\mpl^2$. The simple fact that the horizon crossing happens in two different background dynamics (quasi-de Sitter for inflation and quasi-matter for bounce) changes the tensor-to-scalar ratio by a factor $1/2\epsilon_{c}^{2}$. The factor ${\epsilon_{c}^{2}}/{9}$ has a completely different physical origin. It comes from the dynamics across the bounce.

Inflationary models with adiabatic perturbations have a decreasing and a constant mode. With the quasi-exponential expansion, it is the constant mode that dominates and gives the almost scale-invariant power spectrum. In contrast, bounce models have a constant and an increasing mode before the bounce. The bounce dynamics makes the latter the dominant mode after the bounce (there is a mode mixing), which has an integral contribution of $z^{-2}$ (see Eq.~\eqref{mu_lqc_formal}). This term carries information across the bounce and depends on the dynamics chosen to describe the bounce. In our case we get a ${\epsilon_{c}^2}/{9}$ contribution from the time integral across the LQC bounce. Another bounce like WDW should give a different numerical factor but the same ${\epsilon_{c}^2}$ contribution.

In summary, there is a crucial difference on how inflation and bounce models obtain a small tensor-to-scalar ratio. Both dynamics start with the same vacuum state but the inflationary dynamics amplifies more\footnote{Note that this amplification difference appears only for the tensor-to-scalar ratio and it is irrelevant for the amplitude of the scalar perturbations since one can always adjust the free parameter $\rho_{c}$ in order to accommodate this difference. This is not the case for $r$ since the scalar and tensor perturbations have the same dependence on $\rho_{c}$ hence it drops out (see Eq.s~\eqref{nuout} and \eqref{muout}).} the scalar perturbations than the quasi-matter contraction by a factor of $1/2\epsilon_{c}^2$. On the other hand, the evolution across the bounce suppresses the tensor perturbations by a factor of ${\epsilon_{c}^2}/{9}$. The net result is the $2/9$ difference factor between the two tensor-to-scalar ratios given by Eq.s \eqref{r_lqc}-\eqref{rstab_conc}.

\section{Conclusions}\label{Con}

In the near future, we expect to have decisive new observational data of the very early universe. The 21 cm redshift surveys together with measurements of the CMB B-mode polarization, non-gaussianities and primordial gravitational waves will enable us to discriminate between different primordial universe scenarios. Therefore, it is pressing to identify signatures of each type of primordial universe scenario that would allow us to make testable predictions.

In the present work, we have used Wands' duality to construct a quasi-matter bounce that mimics the Starobinsky inflation. This map allow us to identify the correct contracting phase dynamics that gives the same time-dependent mass term in the Mukhanov-Sasaki equation. The adequate scalar field potential $V(\varphi)$, given by Eq.~\eqref{vphiqmd}, is a deformation of the exponential potential known to describe a pressureless dust fluid. This result agrees with the fact that a quasi-de Sitter phase should be mapped into a quasi-matter dominated contracting universe. After the linear perturbations cross the horizon, the system must go through a bounce phase. We chose to describe the bounce using LQC inasmuch it is the easiest quantum bounce if the matter field is described by a scalar field.

Our constructive method permit us to discriminate the contribution of each dynamical phase in the primordial power spectrum. In particular, we showed that mapping the Starobinsky inflation into a quasi-matter bounce gives the correct relation between the scalar spectral index $n_s-1$ and the tensor-to-scalar ratio $r$ but it appears a factor $2/9$ of difference. The crucial point is to understand the origin of this numerical factor. It comes from the ratio $z_s/z_T$ and it is a convolution of two distinct contribution. The comparison between this ratio from an inflationary expansion to a quasi-matter contraction gives a factor $(2\epsilon_{c}^{2})^{-1}$, while the dynamics through the LQC bounce results in an additional factor of $\epsilon_{c}^{2}/9$.

An interesting feature of our analysis is to show that the bounce leaves a signature in the primordial power spectrum. The scalar and tensor spectral indexes depend on the background dynamics during the horizon crossing. But the amplitudes of the scalar and tensor power spectrum, hence the tensor-to-scalar ratio, carry information from the dynamics across the bounce.

\begin{acknowledgments}
The authors would like to thank and acknowledge financial support from the National Scientific and Technological Research Council (CNPq, Brazil) and the State Scientific and Innovation Funding Agency of Esp\'\i rito Santo (FAPES, Brazil).
\end{acknowledgments}

\bibliography{BIBLIOGRAFIA}

\begin{thebibliography}{72}%
\makeatletter
\providecommand \@ifxundefined [1]{%
 \@ifx{#1\undefined}
}%
\providecommand \@ifnum [1]{%
 \ifnum #1\expandafter \@firstoftwo
 \else \expandafter \@secondoftwo
 \fi
}%
\providecommand \@ifx [1]{%
 \ifx #1\expandafter \@firstoftwo
 \else \expandafter \@secondoftwo
 \fi
}%
\providecommand \natexlab [1]{#1}%
\providecommand \enquote  [1]{``#1''}%
\providecommand \bibnamefont  [1]{#1}%
\providecommand \bibfnamefont [1]{#1}%
\providecommand \citenamefont [1]{#1}%
\providecommand \href@noop [0]{\@secondoftwo}%
\providecommand \href [0]{\begingroup \@sanitize@url \@href}%
\providecommand \@href[1]{\@@startlink{#1}\@@href}%
\providecommand \@@href[1]{\endgroup#1\@@endlink}%
\providecommand \@sanitize@url [0]{\catcode `\\12\catcode `\$12\catcode
  `\&12\catcode `\#12\catcode `\^12\catcode `\_12\catcode `\%12\relax}%
\providecommand \@@startlink[1]{}%
\providecommand \@@endlink[0]{}%
\providecommand \url  [0]{\begingroup\@sanitize@url \@url }%
\providecommand \@url [1]{\endgroup\@href {#1}{\urlprefix }}%
\providecommand \urlprefix  [0]{URL }%
\providecommand \Eprint [0]{\href }%
\providecommand \doibase [0]{http://dx.doi.org/}%
\providecommand \selectlanguage [0]{\@gobble}%
\providecommand \bibinfo  [0]{\@secondoftwo}%
\providecommand \bibfield  [0]{\@secondoftwo}%
\providecommand \translation [1]{[#1]}%
\providecommand \BibitemOpen [0]{}%
\providecommand \bibitemStop [0]{}%
\providecommand \bibitemNoStop [0]{.\EOS\space}%
\providecommand \EOS [0]{\spacefactor3000\relax}%
\providecommand \BibitemShut  [1]{\csname bibitem#1\endcsname}%
\let\auto@bib@innerbib\@empty
\bibitem [{\citenamefont {Aghanim}\ \emph {et~al.}(2018)\citenamefont {Aghanim}
  \emph {et~al.}}]{2018arXiv180706209P}%
  \BibitemOpen
  \bibfield  {author} {\bibinfo {author} {\bibfnamefont {N.}~\bibnamefont
  {Aghanim}} \emph {et~al.},\ }\href@noop {} {\bibfield  {journal} {\bibinfo
  {journal} {arXiv e-prints}\ ,\ \bibinfo {eid} {arXiv:1807.06209}} (\bibinfo
  {year} {2018})},\ \Eprint {http://arxiv.org/abs/1807.06209} {arXiv:1807.06209
  [astro-ph.CO]} \BibitemShut {NoStop}%
\bibitem [{\citenamefont {Akrami}\ \emph
  {et~al.}(2018{\natexlab{a}})\citenamefont {Akrami} \emph
  {et~al.}}]{Akrami:2018odb}%
  \BibitemOpen
  \bibfield  {author} {\bibinfo {author} {\bibfnamefont {Y.}~\bibnamefont
  {Akrami}} \emph {et~al.} (\bibinfo {collaboration} {Planck}),\ }\href@noop {}
  {\  (\bibinfo {year} {2018}{\natexlab{a}})},\ \Eprint
  {http://arxiv.org/abs/1807.06211} {arXiv:1807.06211 [astro-ph.CO]}
  \BibitemShut {NoStop}%
\bibitem [{\citenamefont {Akrami}\ \emph
  {et~al.}(2018{\natexlab{b}})\citenamefont {Akrami} \emph
  {et~al.}}]{2018arXiv180706205P}%
  \BibitemOpen
  \bibfield  {author} {\bibinfo {author} {\bibfnamefont {Y.}~\bibnamefont
  {Akrami}} \emph {et~al.},\ }\href@noop {} {\bibfield  {journal} {\bibinfo
  {journal} {arXiv e-prints}\ ,\ \bibinfo {eid} {arXiv:1807.06205}} (\bibinfo
  {year} {2018}{\natexlab{b}})},\ \Eprint {http://arxiv.org/abs/1807.06205}
  {arXiv:1807.06205 [astro-ph.CO]} \BibitemShut {NoStop}%
\bibitem [{\citenamefont {{Hinshaw}}\ \emph {et~al.}(2013)\citenamefont
  {{Hinshaw}} \emph {et~al.}}]{2013ApJS..208...19H}%
  \BibitemOpen
  \bibfield  {author} {\bibinfo {author} {\bibfnamefont {G.}~\bibnamefont
  {{Hinshaw}}} \emph {et~al.} (\bibinfo {collaboration} {WMAP}),\ }\href
  {\doibase 10.1088/0067-0049/208/2/19} {\bibfield  {journal} {\bibinfo
  {journal} {The Astrophysical Journal Supplement Series}\ }\textbf {\bibinfo
  {volume} {208}},\ \bibinfo {eid} {19} (\bibinfo {year} {2013})},\ \Eprint
  {http://arxiv.org/abs/1212.5226} {arXiv:1212.5226 [astro-ph.CO]} \BibitemShut
  {NoStop}%
\bibitem [{\citenamefont {{Bennett}}\ \emph {et~al.}(2013)\citenamefont
  {{Bennett}} \emph {et~al.}}]{2013ApJS..208...20B}%
  \BibitemOpen
  \bibfield  {author} {\bibinfo {author} {\bibfnamefont {C.~L.}\ \bibnamefont
  {{Bennett}}} \emph {et~al.} (\bibinfo {collaboration} {WMAP}),\ }\href
  {\doibase 10.1088/0067-0049/208/2/20} {\bibfield  {journal} {\bibinfo
  {journal} {The Astrophysical Journal Supplement Series}\ }\textbf {\bibinfo
  {volume} {208}},\ \bibinfo {eid} {20} (\bibinfo {year} {2013})},\ \Eprint
  {http://arxiv.org/abs/1212.5225} {arXiv:1212.5225 [astro-ph.CO]} \BibitemShut
  {NoStop}%
\bibitem [{\citenamefont {Guth}(1981)}]{Guth:1980zm}%
  \BibitemOpen
  \bibfield  {author} {\bibinfo {author} {\bibfnamefont {A.~H.}\ \bibnamefont
  {Guth}},\ }\href {\doibase 10.1103/PhysRevD.23.347} {\bibfield  {journal}
  {\bibinfo  {journal} {Phys. Rev. D}\ }\textbf {\bibinfo {volume} {23}},\
  \bibinfo {pages} {347} (\bibinfo {year} {1981})}\BibitemShut {NoStop}%
\bibitem [{\citenamefont {Linde}(1982)}]{Linde1982389}%
  \BibitemOpen
  \bibfield  {author} {\bibinfo {author} {\bibfnamefont {A.}~\bibnamefont
  {Linde}},\ }\href {\doibase https://doi.org/10.1016/0370-2693(82)91219-9}
  {\bibfield  {journal} {\bibinfo  {journal} {Physics Letters B}\ }\textbf
  {\bibinfo {volume} {108}},\ \bibinfo {pages} {389 } (\bibinfo {year}
  {1982})}\BibitemShut {NoStop}%
\bibitem [{\citenamefont {Albrecht}\ and\ \citenamefont
  {Steinhardt}(1982)}]{Albrecht_inflation}%
  \BibitemOpen
  \bibfield  {author} {\bibinfo {author} {\bibfnamefont {A.}~\bibnamefont
  {Albrecht}}\ and\ \bibinfo {author} {\bibfnamefont {P.~J.}\ \bibnamefont
  {Steinhardt}},\ }\href {\doibase 10.1103/PhysRevLett.48.1220} {\bibfield
  {journal} {\bibinfo  {journal} {Phys. Rev. Lett.}\ }\textbf {\bibinfo
  {volume} {48}},\ \bibinfo {pages} {1220} (\bibinfo {year}
  {1982})}\BibitemShut {NoStop}%
\bibitem [{\citenamefont {Riotto}(2002)}]{Riotto:2002yw}%
  \BibitemOpen
  \bibfield  {author} {\bibinfo {author} {\bibfnamefont {A.}~\bibnamefont
  {Riotto}},\ }in\ \href@noop {} {\emph {\bibinfo {booktitle} {{Astroparticle
  physics and cosmology. Proceedings: Summer School, Trieste, Italy, Jun 17-Jul
  5 2002}}}}\ (\bibinfo {year} {2002})\ pp.\ \bibinfo {pages} {317--413},\
  \Eprint {http://arxiv.org/abs/hep-ph/0210162} {arXiv:hep-ph/0210162 [hep-ph]}
  \BibitemShut {NoStop}%
\bibitem [{\citenamefont {Lyth}\ and\ \citenamefont
  {Riotto}(1999)}]{Lyth:1998xn}%
  \BibitemOpen
  \bibfield  {author} {\bibinfo {author} {\bibfnamefont {D.~H.}\ \bibnamefont
  {Lyth}}\ and\ \bibinfo {author} {\bibfnamefont {A.}~\bibnamefont {Riotto}},\
  }\href {\doibase 10.1016/S0370-1573(98)00128-8} {\bibfield  {journal}
  {\bibinfo  {journal} {Phys. Rept.}\ }\textbf {\bibinfo {volume} {314}},\
  \bibinfo {pages} {1} (\bibinfo {year} {1999})},\ \Eprint
  {http://arxiv.org/abs/hep-ph/9807278} {arXiv:hep-ph/9807278 [hep-ph]}
  \BibitemShut {NoStop}%
\bibitem [{\citenamefont {Brandenberger}(2012)}]{Brandenberger:2012zb}%
  \BibitemOpen
  \bibfield  {author} {\bibinfo {author} {\bibfnamefont {R.~H.}\ \bibnamefont
  {Brandenberger}},\ }\href@noop {} {\  (\bibinfo {year} {2012})},\ \Eprint
  {http://arxiv.org/abs/1206.4196} {arXiv:1206.4196 [astro-ph.CO]} \BibitemShut
  {NoStop}%
\bibitem [{\citenamefont {Falciano}\ \emph
  {et~al.}(2015{\natexlab{a}})\citenamefont {Falciano}, \citenamefont
  {Pinto-Neto},\ and\ \citenamefont {Vitenti}}]{Falciano:2015kya}%
  \BibitemOpen
  \bibfield  {author} {\bibinfo {author} {\bibfnamefont {F.~T.}\ \bibnamefont
  {Falciano}}, \bibinfo {author} {\bibfnamefont {N.}~\bibnamefont
  {Pinto-Neto}}, \ and\ \bibinfo {author} {\bibfnamefont {S.~D.~P.}\
  \bibnamefont {Vitenti}},\ }in\ \href {\doibase 10.1142/9789814623995_0242}
  {\emph {\bibinfo {booktitle} {{Proceedings, 13th Marcel Grossmann Meeting on
  Recent Developments in Theoretical and Experimental General Relativity,
  Astrophysics, and Relativistic Field Theories (MG13): Stockholm, Sweden, July
  1-7, 2012}}}}\ (\bibinfo {year} {2015})\ pp.\ \bibinfo {pages}
  {1625--1627}\BibitemShut {NoStop}%
\bibitem [{\citenamefont {Falciano}\ \emph {et~al.}(2008)\citenamefont
  {Falciano}, \citenamefont {Lilley},\ and\ \citenamefont
  {Peter}}]{Falciano:2008gt}%
  \BibitemOpen
  \bibfield  {author} {\bibinfo {author} {\bibfnamefont {F.~T.}\ \bibnamefont
  {Falciano}}, \bibinfo {author} {\bibfnamefont {M.}~\bibnamefont {Lilley}}, \
  and\ \bibinfo {author} {\bibfnamefont {P.}~\bibnamefont {Peter}},\ }\href
  {\doibase 10.1103/PhysRevD.77.083513} {\bibfield  {journal} {\bibinfo
  {journal} {Phys. Rev.}\ }\textbf {\bibinfo {volume} {D77}},\ \bibinfo {pages}
  {083513} (\bibinfo {year} {2008})},\ \Eprint {http://arxiv.org/abs/0802.1196}
  {arXiv:0802.1196 [gr-qc]} \BibitemShut {NoStop}%
\bibitem [{\citenamefont {Peter}\ and\ \citenamefont
  {Pinto-Neto}(2002)}]{Peter:2002cn}%
  \BibitemOpen
  \bibfield  {author} {\bibinfo {author} {\bibfnamefont {P.}~\bibnamefont
  {Peter}}\ and\ \bibinfo {author} {\bibfnamefont {N.}~\bibnamefont
  {Pinto-Neto}},\ }\href {\doibase 10.1103/PhysRevD.66.063509} {\bibfield
  {journal} {\bibinfo  {journal} {Phys. Rev.}\ }\textbf {\bibinfo {volume}
  {D66}},\ \bibinfo {pages} {063509} (\bibinfo {year} {2002})},\ \Eprint
  {http://arxiv.org/abs/hep-th/0203013} {arXiv:hep-th/0203013 [hep-th]}
  \BibitemShut {NoStop}%
\bibitem [{\citenamefont {Durrer}\ and\ \citenamefont
  {Vernizzi}(2002)}]{Durrer:2002jn}%
  \BibitemOpen
  \bibfield  {author} {\bibinfo {author} {\bibfnamefont {R.}~\bibnamefont
  {Durrer}}\ and\ \bibinfo {author} {\bibfnamefont {F.}~\bibnamefont
  {Vernizzi}},\ }\href {\doibase 10.1103/PhysRevD.66.083503} {\bibfield
  {journal} {\bibinfo  {journal} {Phys. Rev.}\ }\textbf {\bibinfo {volume}
  {D66}},\ \bibinfo {pages} {083503} (\bibinfo {year} {2002})},\ \Eprint
  {http://arxiv.org/abs/hep-ph/0203275} {arXiv:hep-ph/0203275 [hep-ph]}
  \BibitemShut {NoStop}%
\bibitem [{\citenamefont {Enqvist}\ and\ \citenamefont
  {Sloth}(2002)}]{Enqvist:2001zp}%
  \BibitemOpen
  \bibfield  {author} {\bibinfo {author} {\bibfnamefont {K.}~\bibnamefont
  {Enqvist}}\ and\ \bibinfo {author} {\bibfnamefont {M.~S.}\ \bibnamefont
  {Sloth}},\ }\href {\doibase 10.1016/S0550-3213(02)00043-3} {\bibfield
  {journal} {\bibinfo  {journal} {Nucl. Phys.}\ }\textbf {\bibinfo {volume}
  {B626}},\ \bibinfo {pages} {395} (\bibinfo {year} {2002})},\ \Eprint
  {http://arxiv.org/abs/hep-ph/0109214} {arXiv:hep-ph/0109214 [hep-ph]}
  \BibitemShut {NoStop}%
\bibitem [{\citenamefont {Khoury}\ \emph {et~al.}(2002)\citenamefont {Khoury},
  \citenamefont {Ovrut}, \citenamefont {Steinhardt},\ and\ \citenamefont
  {Turok}}]{Khoury:2001zk}%
  \BibitemOpen
  \bibfield  {author} {\bibinfo {author} {\bibfnamefont {J.}~\bibnamefont
  {Khoury}}, \bibinfo {author} {\bibfnamefont {B.~A.}\ \bibnamefont {Ovrut}},
  \bibinfo {author} {\bibfnamefont {P.~J.}\ \bibnamefont {Steinhardt}}, \ and\
  \bibinfo {author} {\bibfnamefont {N.}~\bibnamefont {Turok}},\ }\href
  {\doibase 10.1103/PhysRevD.66.046005} {\bibfield  {journal} {\bibinfo
  {journal} {Phys. Rev.}\ }\textbf {\bibinfo {volume} {D66}},\ \bibinfo {pages}
  {046005} (\bibinfo {year} {2002})},\ \Eprint
  {http://arxiv.org/abs/hep-th/0109050} {arXiv:hep-th/0109050 [hep-th]}
  \BibitemShut {NoStop}%
\bibitem [{\citenamefont {Lyth}(2002)}]{Lyth:2001pf}%
  \BibitemOpen
  \bibfield  {author} {\bibinfo {author} {\bibfnamefont {D.~H.}\ \bibnamefont
  {Lyth}},\ }\href {\doibase 10.1016/S0370-2693(01)01374-0} {\bibfield
  {journal} {\bibinfo  {journal} {Phys. Lett.}\ }\textbf {\bibinfo {volume}
  {B524}},\ \bibinfo {pages} {1} (\bibinfo {year} {2002})},\ \Eprint
  {http://arxiv.org/abs/hep-ph/0106153} {arXiv:hep-ph/0106153 [hep-ph]}
  \BibitemShut {NoStop}%
\bibitem [{\citenamefont {Lehners}\ \emph {et~al.}(2007)\citenamefont
  {Lehners}, \citenamefont {McFadden}, \citenamefont {Turok},\ and\
  \citenamefont {Steinhardt}}]{Lehners:2007ac}%
  \BibitemOpen
  \bibfield  {author} {\bibinfo {author} {\bibfnamefont {J.-L.}\ \bibnamefont
  {Lehners}}, \bibinfo {author} {\bibfnamefont {P.}~\bibnamefont {McFadden}},
  \bibinfo {author} {\bibfnamefont {N.}~\bibnamefont {Turok}}, \ and\ \bibinfo
  {author} {\bibfnamefont {P.~J.}\ \bibnamefont {Steinhardt}},\ }\href
  {\doibase 10.1103/PhysRevD.76.103501} {\bibfield  {journal} {\bibinfo
  {journal} {Phys. Rev.}\ }\textbf {\bibinfo {volume} {D76}},\ \bibinfo {pages}
  {103501} (\bibinfo {year} {2007})},\ \Eprint
  {http://arxiv.org/abs/hep-th/0702153} {arXiv:hep-th/0702153 [HEP-TH]}
  \BibitemShut {NoStop}%
\bibitem [{\citenamefont {Martin}\ \emph {et~al.}(2002)\citenamefont {Martin},
  \citenamefont {Peter}, \citenamefont {Pinto-Neto},\ and\ \citenamefont
  {Schwarz}}]{Martin:2001ue}%
  \BibitemOpen
  \bibfield  {author} {\bibinfo {author} {\bibfnamefont {J.}~\bibnamefont
  {Martin}}, \bibinfo {author} {\bibfnamefont {P.}~\bibnamefont {Peter}},
  \bibinfo {author} {\bibfnamefont {N.}~\bibnamefont {Pinto-Neto}}, \ and\
  \bibinfo {author} {\bibfnamefont {D.~J.}\ \bibnamefont {Schwarz}},\ }\href
  {\doibase 10.1103/PhysRevD.65.123513} {\bibfield  {journal} {\bibinfo
  {journal} {Phys. Rev.}\ }\textbf {\bibinfo {volume} {D65}},\ \bibinfo {pages}
  {123513} (\bibinfo {year} {2002})},\ \Eprint
  {http://arxiv.org/abs/hep-th/0112128} {arXiv:hep-th/0112128 [hep-th]}
  \BibitemShut {NoStop}%
\bibitem [{\citenamefont {Brandenberger}\ and\ \citenamefont
  {Vafa}(1989)}]{Brandenberger1989391}%
  \BibitemOpen
  \bibfield  {author} {\bibinfo {author} {\bibfnamefont {R.}~\bibnamefont
  {Brandenberger}}\ and\ \bibinfo {author} {\bibfnamefont {C.}~\bibnamefont
  {Vafa}},\ }\href {\doibase https://doi.org/10.1016/0550-3213(89)90037-0}
  {\bibfield  {journal} {\bibinfo  {journal} {Nuclear Physics B}\ }\textbf
  {\bibinfo {volume} {316}},\ \bibinfo {pages} {391 } (\bibinfo {year}
  {1989})}\BibitemShut {NoStop}%
\bibitem [{\citenamefont {Battefeld}\ and\ \citenamefont
  {Watson}(2006)}]{String_gas_Battefeld}%
  \BibitemOpen
  \bibfield  {author} {\bibinfo {author} {\bibfnamefont {T.}~\bibnamefont
  {Battefeld}}\ and\ \bibinfo {author} {\bibfnamefont {S.}~\bibnamefont
  {Watson}},\ }\href {\doibase 10.1103/RevModPhys.78.435} {\bibfield  {journal}
  {\bibinfo  {journal} {Rev. Mod. Phys.}\ }\textbf {\bibinfo {volume} {78}},\
  \bibinfo {pages} {435} (\bibinfo {year} {2006})}\BibitemShut {NoStop}%
\bibitem [{\citenamefont {Matsumura}\ \emph {et~al.}(2013)\citenamefont
  {Matsumura} \emph {et~al.}}]{Matsumura:2013aja}%
  \BibitemOpen
  \bibfield  {author} {\bibinfo {author} {\bibfnamefont {T.}~\bibnamefont
  {Matsumura}} \emph {et~al.},\ }\href {\doibase 10.1007/s10909-013-0996-1} {\
  (\bibinfo {year} {2013}),\ 10.1007/s10909-013-0996-1},\ \bibinfo {note} {[J.
  Low. Temp. Phys.176,733(2014)]},\ \Eprint {http://arxiv.org/abs/1311.2847}
  {arXiv:1311.2847 [astro-ph.IM]} \BibitemShut {NoStop}%
\bibitem [{\citenamefont {Abazajian}\ \emph {et~al.}(2016)\citenamefont
  {Abazajian} \emph {et~al.}}]{Abazajian:2016yjj}%
  \BibitemOpen
  \bibfield  {author} {\bibinfo {author} {\bibfnamefont {K.~N.}\ \bibnamefont
  {Abazajian}} \emph {et~al.} (\bibinfo {collaboration} {CMB-S4}),\ }\href@noop
  {} {\  (\bibinfo {year} {2016})},\ \Eprint {http://arxiv.org/abs/1610.02743}
  {arXiv:1610.02743 [astro-ph.CO]} \BibitemShut {NoStop}%
\bibitem [{\citenamefont {Collaboration}(2018)}]{Ade:2018sbj}%
  \BibitemOpen
  \bibfield  {author} {\bibinfo {author} {\bibfnamefont {T.~S.~O.}\
  \bibnamefont {Collaboration}} (\bibinfo {collaboration} {Simons
  Observatory}),\ }\href@noop {} {\  (\bibinfo {year} {2018})},\ \Eprint
  {http://arxiv.org/abs/1808.07445} {arXiv:1808.07445 [astro-ph.CO]}
  \BibitemShut {NoStop}%
\bibitem [{\citenamefont {collaboration}(2011)}]{Bouchet:2011ck}%
  \BibitemOpen
  \bibfield  {author} {\bibinfo {author} {\bibfnamefont {T.~C.}\ \bibnamefont
  {collaboration}} (\bibinfo {collaboration} {COrE}),\ }\href@noop {} {\
  (\bibinfo {year} {2011})},\ \Eprint {http://arxiv.org/abs/1102.2181}
  {arXiv:1102.2181 [astro-ph.CO]} \BibitemShut {NoStop}%
\bibitem [{\citenamefont {Wands}(1999)}]{Wands:1998yp}%
  \BibitemOpen
  \bibfield  {author} {\bibinfo {author} {\bibfnamefont {D.}~\bibnamefont
  {Wands}},\ }\href {\doibase 10.1103/PhysRevD.60.023507} {\bibfield  {journal}
  {\bibinfo  {journal} {Phys. Rev.}\ }\textbf {\bibinfo {volume} {D60}},\
  \bibinfo {pages} {023507} (\bibinfo {year} {1999})},\ \Eprint
  {http://arxiv.org/abs/gr-qc/9809062} {arXiv:gr-qc/9809062 [gr-qc]}
  \BibitemShut {NoStop}%
\bibitem [{\citenamefont {Brustein}\ \emph {et~al.}(1998)\citenamefont
  {Brustein}, \citenamefont {Gasperini},\ and\ \citenamefont
  {Veneziano}}]{Brustein:1998kq}%
  \BibitemOpen
  \bibfield  {author} {\bibinfo {author} {\bibfnamefont {R.}~\bibnamefont
  {Brustein}}, \bibinfo {author} {\bibfnamefont {M.}~\bibnamefont {Gasperini}},
  \ and\ \bibinfo {author} {\bibfnamefont {G.}~\bibnamefont {Veneziano}},\
  }\href {\doibase 10.1016/S0370-2693(98)00576-0} {\bibfield  {journal}
  {\bibinfo  {journal} {Phys. Lett.}\ }\textbf {\bibinfo {volume} {B431}},\
  \bibinfo {pages} {277} (\bibinfo {year} {1998})},\ \Eprint
  {http://arxiv.org/abs/hep-th/9803018} {arXiv:hep-th/9803018 [hep-th]}
  \BibitemShut {NoStop}%
\bibitem [{\citenamefont {Finelli}\ and\ \citenamefont
  {Brandenberger}(2002)}]{Finelli:2001sr}%
  \BibitemOpen
  \bibfield  {author} {\bibinfo {author} {\bibfnamefont {F.}~\bibnamefont
  {Finelli}}\ and\ \bibinfo {author} {\bibfnamefont {R.}~\bibnamefont
  {Brandenberger}},\ }\href {\doibase 10.1103/PhysRevD.65.103522} {\bibfield
  {journal} {\bibinfo  {journal} {Phys. Rev.}\ }\textbf {\bibinfo {volume}
  {D65}},\ \bibinfo {pages} {103522} (\bibinfo {year} {2002})},\ \Eprint
  {http://arxiv.org/abs/hep-th/0112249} {arXiv:hep-th/0112249 [hep-th]}
  \BibitemShut {NoStop}%
\bibitem [{\citenamefont {Mukhanov}\ \emph {et~al.}(1992)\citenamefont
  {Mukhanov}, \citenamefont {Feldman},\ and\ \citenamefont
  {Brandenberger}}]{Mukhanov:1990me}%
  \BibitemOpen
  \bibfield  {author} {\bibinfo {author} {\bibfnamefont {V.~F.}\ \bibnamefont
  {Mukhanov}}, \bibinfo {author} {\bibfnamefont {H.~A.}\ \bibnamefont
  {Feldman}}, \ and\ \bibinfo {author} {\bibfnamefont {R.~H.}\ \bibnamefont
  {Brandenberger}},\ }\href {\doibase 10.1016/0370-1573(92)90044-Z} {\bibfield
  {journal} {\bibinfo  {journal} {Phys. Rept.}\ }\textbf {\bibinfo {volume}
  {215}},\ \bibinfo {pages} {203} (\bibinfo {year} {1992})}\BibitemShut
  {NoStop}%
\bibitem [{\citenamefont
  {{Brandenberger}}(2004)}]{Brandenberger:2004LNP...646..127B}%
  \BibitemOpen
  \bibfield  {author} {\bibinfo {author} {\bibfnamefont {R.~H.}\ \bibnamefont
  {{Brandenberger}}},\ }\enquote {\bibinfo {title} {{Lectures on the Theory of
  Cosmological Perturbations}},}\ in\ \href {\doibase
  10.1007/978-3-540-40918-2_5} {\emph {\bibinfo {booktitle} {The early universe
  and observational cosmology, edited by N. Breton, J.L. Cervantes-Cota, and M.
  Salgado, Lecture Notes in Physics, vol. 646, 2004., p.127-167}}},\ \bibinfo
  {editor} {edited by\ \bibinfo {editor} {\bibfnamefont {N.}~\bibnamefont
  {{Bret{\'o}n}}}, \bibinfo {editor} {\bibfnamefont {J.~L.}\ \bibnamefont
  {{Cervantes-Cota}}}, \ and\ \bibinfo {editor} {\bibfnamefont
  {M.}~\bibnamefont {{Salgad}}}}\ (\bibinfo {year} {2004})\ pp.\ \bibinfo
  {pages} {127--167}\BibitemShut {NoStop}%
\bibitem [{\citenamefont {Starobinsky}(1980)}]{Starobinsky:1980te}%
  \BibitemOpen
  \bibfield  {author} {\bibinfo {author} {\bibfnamefont {A.~A.}\ \bibnamefont
  {Starobinsky}},\ }\href {\doibase 10.1016/0370-2693(80)90670-X} {\bibfield
  {journal} {\bibinfo  {journal} {Phys. Lett.}\ }\textbf {\bibinfo {volume}
  {B91}},\ \bibinfo {pages} {99} (\bibinfo {year} {1980})}\BibitemShut
  {NoStop}%
\bibitem [{\citenamefont {De~Felice}\ and\ \citenamefont
  {Tsujikawa}(2010)}]{DeFelice:2010aj}%
  \BibitemOpen
  \bibfield  {author} {\bibinfo {author} {\bibfnamefont {A.}~\bibnamefont
  {De~Felice}}\ and\ \bibinfo {author} {\bibfnamefont {S.}~\bibnamefont
  {Tsujikawa}},\ }\href {\doibase 10.12942/lrr-2010-3} {\bibfield  {journal}
  {\bibinfo  {journal} {Living Rev. Rel.}\ }\textbf {\bibinfo {volume} {13}},\
  \bibinfo {pages} {3} (\bibinfo {year} {2010})},\ \Eprint
  {http://arxiv.org/abs/1002.4928} {arXiv:1002.4928 [gr-qc]} \BibitemShut
  {NoStop}%
\bibitem [{\citenamefont {Hwang}(1997)}]{Hwang:1996bc}%
  \BibitemOpen
  \bibfield  {author} {\bibinfo {author} {\bibfnamefont {J.-c.}\ \bibnamefont
  {Hwang}},\ }\href {\doibase 10.1088/0264-9381/14/12/016} {\bibfield
  {journal} {\bibinfo  {journal} {Class. Quant. Grav.}\ }\textbf {\bibinfo
  {volume} {14}},\ \bibinfo {pages} {3327} (\bibinfo {year} {1997})},\ \Eprint
  {http://arxiv.org/abs/gr-qc/9607059} {arXiv:gr-qc/9607059 [gr-qc]}
  \BibitemShut {NoStop}%
\bibitem [{\citenamefont {Hwang}\ and\ \citenamefont
  {Noh}(1996)}]{Hwang:1996xh}%
  \BibitemOpen
  \bibfield  {author} {\bibinfo {author} {\bibfnamefont {J.-c.}\ \bibnamefont
  {Hwang}}\ and\ \bibinfo {author} {\bibfnamefont {H.}~\bibnamefont {Noh}},\
  }\href {\doibase 10.1103/PhysRevD.54.1460} {\bibfield  {journal} {\bibinfo
  {journal} {Phys. Rev.}\ }\textbf {\bibinfo {volume} {D54}},\ \bibinfo {pages}
  {1460} (\bibinfo {year} {1996})}\BibitemShut {NoStop}%
\bibitem [{\citenamefont {Sotiriou}\ and\ \citenamefont
  {Faraoni}(2010)}]{Sotiriou:2008rp}%
  \BibitemOpen
  \bibfield  {author} {\bibinfo {author} {\bibfnamefont {T.~P.}\ \bibnamefont
  {Sotiriou}}\ and\ \bibinfo {author} {\bibfnamefont {V.}~\bibnamefont
  {Faraoni}},\ }\href {\doibase 10.1103/RevModPhys.82.451} {\bibfield
  {journal} {\bibinfo  {journal} {Rev. Mod. Phys.}\ }\textbf {\bibinfo {volume}
  {82}},\ \bibinfo {pages} {451} (\bibinfo {year} {2010})},\ \Eprint
  {http://arxiv.org/abs/0805.1726} {arXiv:0805.1726 [gr-qc]} \BibitemShut
  {NoStop}%
\bibitem [{\citenamefont {Mukhanov}(2005)}]{Mukhanov:2005sc}%
  \BibitemOpen
  \bibfield  {author} {\bibinfo {author} {\bibfnamefont {V.}~\bibnamefont
  {Mukhanov}},\ }\href
  {http://www-spires.fnal.gov/spires/find/books/www?cl=QB981.M89::2005} {\emph
  {\bibinfo {title} {{Physical Foundations of Cosmology}}}}\ (\bibinfo
  {publisher} {Cambridge University Press},\ \bibinfo {address} {Oxford},\
  \bibinfo {year} {2005})\BibitemShut {NoStop}%
\bibitem [{\citenamefont {Kamenshchik}\ \emph {et~al.}(2016)\citenamefont
  {Kamenshchik}, \citenamefont {Pozdeeva}, \citenamefont {Vernov},
  \citenamefont {Tronconi},\ and\ \citenamefont
  {Venturi}}]{Kamenshchik:2016gcy}%
  \BibitemOpen
  \bibfield  {author} {\bibinfo {author} {\bibfnamefont {A.~{\relax Yu}.}\
  \bibnamefont {Kamenshchik}}, \bibinfo {author} {\bibfnamefont {E.~O.}\
  \bibnamefont {Pozdeeva}}, \bibinfo {author} {\bibfnamefont {S.~{\relax Yu}.}\
  \bibnamefont {Vernov}}, \bibinfo {author} {\bibfnamefont {A.}~\bibnamefont
  {Tronconi}}, \ and\ \bibinfo {author} {\bibfnamefont {G.}~\bibnamefont
  {Venturi}},\ }\href {\doibase 10.1103/PhysRevD.94.063510} {\bibfield
  {journal} {\bibinfo  {journal} {Phys. Rev.}\ }\textbf {\bibinfo {volume}
  {D94}},\ \bibinfo {pages} {063510} (\bibinfo {year} {2016})},\ \Eprint
  {http://arxiv.org/abs/1602.07192} {arXiv:1602.07192 [gr-qc]} \BibitemShut
  {NoStop}%
\bibitem [{\citenamefont {Koshelev}\ \emph {et~al.}(2016)\citenamefont
  {Koshelev}, \citenamefont {Modesto}, \citenamefont {Rachwal},\ and\
  \citenamefont {Starobinsky}}]{Koshelev:2016xqb}%
  \BibitemOpen
  \bibfield  {author} {\bibinfo {author} {\bibfnamefont {A.~S.}\ \bibnamefont
  {Koshelev}}, \bibinfo {author} {\bibfnamefont {L.}~\bibnamefont {Modesto}},
  \bibinfo {author} {\bibfnamefont {L.}~\bibnamefont {Rachwal}}, \ and\
  \bibinfo {author} {\bibfnamefont {A.~A.}\ \bibnamefont {Starobinsky}},\
  }\href {\doibase 10.1007/JHEP11(2016)067} {\bibfield  {journal} {\bibinfo
  {journal} {JHEP}\ }\textbf {\bibinfo {volume} {11}},\ \bibinfo {pages} {067}
  (\bibinfo {year} {2016})},\ \Eprint {http://arxiv.org/abs/1604.03127}
  {arXiv:1604.03127 [hep-th]} \BibitemShut {NoStop}%
\bibitem [{\citenamefont {Wilson-Ewing}(2013)}]{WilsonEwing:2012pu}%
  \BibitemOpen
  \bibfield  {author} {\bibinfo {author} {\bibfnamefont {E.}~\bibnamefont
  {Wilson-Ewing}},\ }\href {\doibase 10.1088/1475-7516/2013/03/026} {\bibfield
  {journal} {\bibinfo  {journal} {JCAP}\ }\textbf {\bibinfo {volume} {1303}},\
  \bibinfo {pages} {026} (\bibinfo {year} {2013})},\ \Eprint
  {http://arxiv.org/abs/1211.6269} {arXiv:1211.6269 [gr-qc]} \BibitemShut
  {NoStop}%
\bibitem [{\citenamefont {de~Haro}\ and\ \citenamefont
  {Cai}(2015)}]{deHaro:2015wda}%
  \BibitemOpen
  \bibfield  {author} {\bibinfo {author} {\bibfnamefont {J.}~\bibnamefont
  {de~Haro}}\ and\ \bibinfo {author} {\bibfnamefont {Y.-F.}\ \bibnamefont
  {Cai}},\ }\href {\doibase 10.1007/s10714-015-1936-y} {\bibfield  {journal}
  {\bibinfo  {journal} {Gen. Rel. Grav.}\ }\textbf {\bibinfo {volume} {47}},\
  \bibinfo {pages} {95} (\bibinfo {year} {2015})},\ \Eprint
  {http://arxiv.org/abs/1502.03230} {arXiv:1502.03230 [gr-qc]} \BibitemShut
  {NoStop}%
\bibitem [{\citenamefont {Borde}\ and\ \citenamefont
  {Vilenkin}(1994)}]{Borde:1994PRL72}%
  \BibitemOpen
  \bibfield  {author} {\bibinfo {author} {\bibfnamefont {A.}~\bibnamefont
  {Borde}}\ and\ \bibinfo {author} {\bibfnamefont {A.}~\bibnamefont
  {Vilenkin}},\ }\href {\doibase 10.1103/PhysRevLett.72.3305} {\bibfield
  {journal} {\bibinfo  {journal} {Phys. Rev. Lett.}\ }\textbf {\bibinfo
  {volume} {72}},\ \bibinfo {pages} {3305} (\bibinfo {year}
  {1994})}\BibitemShut {NoStop}%
\bibitem [{\citenamefont {BORDE}\ and\ \citenamefont
  {VILENKIN}(1996)}]{Borde_IJMP1996}%
  \BibitemOpen
  \bibfield  {author} {\bibinfo {author} {\bibfnamefont {A.}~\bibnamefont
  {BORDE}}\ and\ \bibinfo {author} {\bibfnamefont {A.}~\bibnamefont
  {VILENKIN}},\ }\href {\doibase 10.1142/S0218271896000497} {\bibfield
  {journal} {\bibinfo  {journal} {International Journal of Modern Physics D}\
  }\textbf {\bibinfo {volume} {05}},\ \bibinfo {pages} {813} (\bibinfo {year}
  {1996})},\ \Eprint
  {http://arxiv.org/abs/https://doi.org/10.1142/S0218271896000497}
  {https://doi.org/10.1142/S0218271896000497} \BibitemShut {NoStop}%
\bibitem [{\citenamefont {Falciano}\ \emph {et~al.}(2007)\citenamefont
  {Falciano}, \citenamefont {Pinto-Neto},\ and\ \citenamefont
  {Santini}}]{Falciano:2007yf}%
  \BibitemOpen
  \bibfield  {author} {\bibinfo {author} {\bibfnamefont {F.~T.}\ \bibnamefont
  {Falciano}}, \bibinfo {author} {\bibfnamefont {N.}~\bibnamefont
  {Pinto-Neto}}, \ and\ \bibinfo {author} {\bibfnamefont {E.~S.}\ \bibnamefont
  {Santini}},\ }\href {\doibase 10.1103/PhysRevD.76.083521} {\bibfield
  {journal} {\bibinfo  {journal} {Phys. Rev.}\ }\textbf {\bibinfo {volume}
  {D76}},\ \bibinfo {pages} {083521} (\bibinfo {year} {2007})},\ \Eprint
  {http://arxiv.org/abs/0707.1088} {arXiv:0707.1088 [gr-qc]} \BibitemShut
  {NoStop}%
\bibitem [{\citenamefont {Falciano}\ and\ \citenamefont
  {Pinto-Neto}(2009)}]{Falciano:2008nk}%
  \BibitemOpen
  \bibfield  {author} {\bibinfo {author} {\bibfnamefont {F.~T.}\ \bibnamefont
  {Falciano}}\ and\ \bibinfo {author} {\bibfnamefont {N.}~\bibnamefont
  {Pinto-Neto}},\ }\href {\doibase 10.1103/PhysRevD.79.023507} {\bibfield
  {journal} {\bibinfo  {journal} {Phys. Rev.}\ }\textbf {\bibinfo {volume}
  {D79}},\ \bibinfo {pages} {023507} (\bibinfo {year} {2009})},\ \Eprint
  {http://arxiv.org/abs/0810.3542} {arXiv:0810.3542 [gr-qc]} \BibitemShut
  {NoStop}%
\bibitem [{\citenamefont {Peter}\ \emph {et~al.}(2007)\citenamefont {Peter},
  \citenamefont {Pinho},\ and\ \citenamefont {Pinto-Neto}}]{NPN_Peter_Pinho_1}%
  \BibitemOpen
  \bibfield  {author} {\bibinfo {author} {\bibfnamefont {P.}~\bibnamefont
  {Peter}}, \bibinfo {author} {\bibfnamefont {E.~J.~C.}\ \bibnamefont {Pinho}},
  \ and\ \bibinfo {author} {\bibfnamefont {N.}~\bibnamefont {Pinto-Neto}},\
  }\href {\doibase 10.1103/PhysRevD.75.023516} {\bibfield  {journal} {\bibinfo
  {journal} {Phys. Rev. D}\ }\textbf {\bibinfo {volume} {75}},\ \bibinfo
  {pages} {023516} (\bibinfo {year} {2007})}\BibitemShut {NoStop}%
\bibitem [{\citenamefont {Peter}\ and\ \citenamefont
  {Pinto-Neto}(2008)}]{Peter:2008qz}%
  \BibitemOpen
  \bibfield  {author} {\bibinfo {author} {\bibfnamefont {P.}~\bibnamefont
  {Peter}}\ and\ \bibinfo {author} {\bibfnamefont {N.}~\bibnamefont
  {Pinto-Neto}},\ }\href {\doibase 10.1103/PhysRevD.78.063506} {\bibfield
  {journal} {\bibinfo  {journal} {Phys. Rev.}\ }\textbf {\bibinfo {volume}
  {D78}},\ \bibinfo {pages} {063506} (\bibinfo {year} {2008})},\ \Eprint
  {http://arxiv.org/abs/0809.2022} {arXiv:0809.2022 [gr-qc]} \BibitemShut
  {NoStop}%
\bibitem [{\citenamefont {Bojowald}(2015)}]{Bojowald_RepProgPhys2015}%
  \BibitemOpen
  \bibfield  {author} {\bibinfo {author} {\bibfnamefont {M.}~\bibnamefont
  {Bojowald}},\ }\href {http://stacks.iop.org/0034-4885/78/i=2/a=023901}
  {\bibfield  {journal} {\bibinfo  {journal} {Reports on Progress in Physics}\
  }\textbf {\bibinfo {volume} {78}},\ \bibinfo {pages} {023901} (\bibinfo
  {year} {2015})}\BibitemShut {NoStop}%
\bibitem [{\citenamefont {Cai}\ and\ \citenamefont
  {Wilson-Ewing}(2014)}]{Cai:2014zga}%
  \BibitemOpen
  \bibfield  {author} {\bibinfo {author} {\bibfnamefont {Y.-F.}\ \bibnamefont
  {Cai}}\ and\ \bibinfo {author} {\bibfnamefont {E.}~\bibnamefont
  {Wilson-Ewing}},\ }\href {\doibase 10.1088/1475-7516/2014/03/026} {\bibfield
  {journal} {\bibinfo  {journal} {JCAP}\ }\textbf {\bibinfo {volume} {1403}},\
  \bibinfo {pages} {026} (\bibinfo {year} {2014})},\ \Eprint
  {http://arxiv.org/abs/1402.3009} {arXiv:1402.3009 [gr-qc]} \BibitemShut
  {NoStop}%
\bibitem [{\citenamefont {Craig}\ and\ \citenamefont
  {Singh}(2013)}]{Craig_Singh_CQG2013}%
  \BibitemOpen
  \bibfield  {author} {\bibinfo {author} {\bibfnamefont {D.~A.}\ \bibnamefont
  {Craig}}\ and\ \bibinfo {author} {\bibfnamefont {P.}~\bibnamefont {Singh}},\
  }\href {http://stacks.iop.org/0264-9381/30/i=20/a=205008} {\bibfield
  {journal} {\bibinfo  {journal} {Classical and Quantum Gravity}\ }\textbf
  {\bibinfo {volume} {30}},\ \bibinfo {pages} {205008} (\bibinfo {year}
  {2013})}\BibitemShut {NoStop}%
\bibitem [{\citenamefont {Ashtekar}\ \emph
  {et~al.}(2006{\natexlab{a}})\citenamefont {Ashtekar}, \citenamefont
  {Pawlowski},\ and\ \citenamefont {Singh}}]{Astekar_PRD74}%
  \BibitemOpen
  \bibfield  {author} {\bibinfo {author} {\bibfnamefont {A.}~\bibnamefont
  {Ashtekar}}, \bibinfo {author} {\bibfnamefont {T.}~\bibnamefont {Pawlowski}},
  \ and\ \bibinfo {author} {\bibfnamefont {P.}~\bibnamefont {Singh}},\ }\href
  {\doibase 10.1103/PhysRevD.74.084003} {\bibfield  {journal} {\bibinfo
  {journal} {Phys. Rev. D}\ }\textbf {\bibinfo {volume} {74}},\ \bibinfo
  {pages} {084003} (\bibinfo {year} {2006}{\natexlab{a}})}\BibitemShut
  {NoStop}%
\bibitem [{\citenamefont {Singh}\ \emph {et~al.}(2006)\citenamefont {Singh},
  \citenamefont {Vandersloot},\ and\ \citenamefont
  {Vereshchagin}}]{Singh_PRD.74.2006}%
  \BibitemOpen
  \bibfield  {author} {\bibinfo {author} {\bibfnamefont {P.}~\bibnamefont
  {Singh}}, \bibinfo {author} {\bibfnamefont {K.}~\bibnamefont {Vandersloot}},
  \ and\ \bibinfo {author} {\bibfnamefont {G.~V.}\ \bibnamefont
  {Vereshchagin}},\ }\href {\doibase 10.1103/PhysRevD.74.043510} {\bibfield
  {journal} {\bibinfo  {journal} {Phys. Rev. D}\ }\textbf {\bibinfo {volume}
  {74}},\ \bibinfo {pages} {043510} (\bibinfo {year} {2006})}\BibitemShut
  {NoStop}%
\bibitem [{\citenamefont {Brandenberger}(2000)}]{Brandenberger:2000as}%
  \BibitemOpen
  \bibfield  {author} {\bibinfo {author} {\bibfnamefont {R.~H.}\ \bibnamefont
  {Brandenberger}},\ }in\ \href@noop {} {\emph {\bibinfo {booktitle}
  {{Proceedings, 10th Workshop on General Relativity and Gravitation in Japan
  (JGRG10): Osaka, Japan, September 10-14, 2000}}}}\ (\bibinfo {year} {2000})\
  \Eprint {http://arxiv.org/abs/hep-ph/0101119} {arXiv:hep-ph/0101119 [hep-ph]}
  \BibitemShut {NoStop}%
\bibitem [{\citenamefont {Martin}\ and\ \citenamefont
  {Brandenberger}(2001)}]{Jerome_PhysRevD.63.123501}%
  \BibitemOpen
  \bibfield  {author} {\bibinfo {author} {\bibfnamefont {J.}~\bibnamefont
  {Martin}}\ and\ \bibinfo {author} {\bibfnamefont {R.~H.}\ \bibnamefont
  {Brandenberger}},\ }\href {\doibase 10.1103/PhysRevD.63.123501} {\bibfield
  {journal} {\bibinfo  {journal} {Phys. Rev. D}\ }\textbf {\bibinfo {volume}
  {63}},\ \bibinfo {pages} {123501} (\bibinfo {year} {2001})}\BibitemShut
  {NoStop}%
\bibitem [{\citenamefont {Battefeld}\ and\ \citenamefont
  {Peter}(2015)}]{Battefeld:2014uga}%
  \BibitemOpen
  \bibfield  {author} {\bibinfo {author} {\bibfnamefont {D.}~\bibnamefont
  {Battefeld}}\ and\ \bibinfo {author} {\bibfnamefont {P.}~\bibnamefont
  {Peter}},\ }\href {\doibase 10.1016/j.physrep.2014.12.004} {\bibfield
  {journal} {\bibinfo  {journal} {Phys. Rept.}\ }\textbf {\bibinfo {volume}
  {571}},\ \bibinfo {pages} {1} (\bibinfo {year} {2015})},\ \Eprint
  {http://arxiv.org/abs/1406.2790} {arXiv:1406.2790 [astro-ph.CO]} \BibitemShut
  {NoStop}%
\bibitem [{\citenamefont
  {Brandenberger}(2011{\natexlab{a}})}]{Brandenberger:2009jq}%
  \BibitemOpen
  \bibfield  {author} {\bibinfo {author} {\bibfnamefont {R.~H.}\ \bibnamefont
  {Brandenberger}},\ }\bibfield  {booktitle} {\emph {\bibinfo {booktitle}
  {{Proceedings, 5th International Symposium on Cosmology and Particle
  Astrophysics (CosPA 2008): Pohang, Korea, October 29-November 1, 2008}}},\
  }\href {\doibase 10.1142/S2010194511000109} {\bibfield  {journal} {\bibinfo
  {journal} {Int. J. Mod. Phys. Conf. Ser.}\ }\textbf {\bibinfo {volume}
  {01}},\ \bibinfo {pages} {67} (\bibinfo {year} {2011}{\natexlab{a}})},\
  \Eprint {http://arxiv.org/abs/0902.4731} {arXiv:0902.4731 [hep-th]}
  \BibitemShut {NoStop}%
\bibitem [{\citenamefont {Weinberg}(2003)}]{Weinberg:2003sw}%
  \BibitemOpen
  \bibfield  {author} {\bibinfo {author} {\bibfnamefont {S.}~\bibnamefont
  {Weinberg}},\ }\href {\doibase 10.1103/PhysRevD.67.123504} {\bibfield
  {journal} {\bibinfo  {journal} {Phys. Rev.}\ }\textbf {\bibinfo {volume}
  {D67}},\ \bibinfo {pages} {123504} (\bibinfo {year} {2003})},\ \Eprint
  {http://arxiv.org/abs/astro-ph/0302326} {arXiv:astro-ph/0302326 [astro-ph]}
  \BibitemShut {NoStop}%
\bibitem [{\citenamefont {Cai}\ and\ \citenamefont
  {Wilson-Ewing}(2015)}]{WilsonEwingLCDM}%
  \BibitemOpen
  \bibfield  {author} {\bibinfo {author} {\bibfnamefont {Y.-F.}\ \bibnamefont
  {Cai}}\ and\ \bibinfo {author} {\bibfnamefont {E.}~\bibnamefont
  {Wilson-Ewing}},\ }\href {http://stacks.iop.org/1475-7516/2015/i=03/a=006}
  {\bibfield  {journal} {\bibinfo  {journal} {Journal of Cosmology and
  Astroparticle Physics}\ }\textbf {\bibinfo {volume} {2015}},\ \bibinfo
  {pages} {006} (\bibinfo {year} {2015})}\BibitemShut {NoStop}%
\bibitem [{\citenamefont {Pinto-Neto}\ and\ \citenamefont
  {Fabris}(2013)}]{Pinto-Neto:2013toa}%
  \BibitemOpen
  \bibfield  {author} {\bibinfo {author} {\bibfnamefont {N.}~\bibnamefont
  {Pinto-Neto}}\ and\ \bibinfo {author} {\bibfnamefont {J.~C.}\ \bibnamefont
  {Fabris}},\ }\href {\doibase 10.1088/0264-9381/30/14/143001} {\bibfield
  {journal} {\bibinfo  {journal} {Class. Quant. Grav.}\ }\textbf {\bibinfo
  {volume} {30}},\ \bibinfo {pages} {143001} (\bibinfo {year} {2013})},\
  \Eprint {http://arxiv.org/abs/1306.0820} {arXiv:1306.0820 [gr-qc]}
  \BibitemShut {NoStop}%
\bibitem [{\citenamefont {Pinto-Neto}\ \emph {et~al.}(2012)\citenamefont
  {Pinto-Neto}, \citenamefont {Falciano}, \citenamefont {Pereira},\ and\
  \citenamefont {Santini}}]{PintoNeto:2012ug}%
  \BibitemOpen
  \bibfield  {author} {\bibinfo {author} {\bibfnamefont {N.}~\bibnamefont
  {Pinto-Neto}}, \bibinfo {author} {\bibfnamefont {F.~T.}\ \bibnamefont
  {Falciano}}, \bibinfo {author} {\bibfnamefont {R.}~\bibnamefont {Pereira}}, \
  and\ \bibinfo {author} {\bibfnamefont {E.~S.}\ \bibnamefont {Santini}},\
  }\href {\doibase 10.1103/PhysRevD.86.063504} {\bibfield  {journal} {\bibinfo
  {journal} {Phys. Rev.}\ }\textbf {\bibinfo {volume} {D86}},\ \bibinfo {pages}
  {063504} (\bibinfo {year} {2012})},\ \Eprint {http://arxiv.org/abs/1206.4021}
  {arXiv:1206.4021 [gr-qc]} \BibitemShut {NoStop}%
\bibitem [{\citenamefont {Vitenti}\ \emph {et~al.}(2013)\citenamefont
  {Vitenti}, \citenamefont {Falciano},\ and\ \citenamefont
  {Pinto-Neto}}]{Vitenti:2012cx}%
  \BibitemOpen
  \bibfield  {author} {\bibinfo {author} {\bibfnamefont {S.~D.~P.}\
  \bibnamefont {Vitenti}}, \bibinfo {author} {\bibfnamefont {F.~T.}\
  \bibnamefont {Falciano}}, \ and\ \bibinfo {author} {\bibfnamefont
  {N.}~\bibnamefont {Pinto-Neto}},\ }\href {\doibase
  10.1103/PhysRevD.87.103503} {\bibfield  {journal} {\bibinfo  {journal} {Phys.
  Rev.}\ }\textbf {\bibinfo {volume} {D87}},\ \bibinfo {pages} {103503}
  (\bibinfo {year} {2013})},\ \Eprint {http://arxiv.org/abs/1206.4374}
  {arXiv:1206.4374 [gr-qc]} \BibitemShut {NoStop}%
\bibitem [{\citenamefont {Falciano}\ \emph
  {et~al.}(2015{\natexlab{b}})\citenamefont {Falciano}, \citenamefont
  {Pinto-Neto},\ and\ \citenamefont {Struyve}}]{Falciano_PRD_2015}%
  \BibitemOpen
  \bibfield  {author} {\bibinfo {author} {\bibfnamefont {F.~T.}\ \bibnamefont
  {Falciano}}, \bibinfo {author} {\bibfnamefont {N.}~\bibnamefont
  {Pinto-Neto}}, \ and\ \bibinfo {author} {\bibfnamefont {W.}~\bibnamefont
  {Struyve}},\ }\href {\doibase 10.1103/PhysRevD.91.043524} {\bibfield
  {journal} {\bibinfo  {journal} {Phys. Rev. D}\ }\textbf {\bibinfo {volume}
  {91}},\ \bibinfo {pages} {043524} (\bibinfo {year}
  {2015}{\natexlab{b}})}\BibitemShut {NoStop}%
\bibitem [{\citenamefont {Pinto-Neto}\ \emph {et~al.}(2005)\citenamefont
  {Pinto-Neto}, \citenamefont {Santini},\ and\ \citenamefont
  {Falciano}}]{PintoNeto:2005gx}%
  \BibitemOpen
  \bibfield  {author} {\bibinfo {author} {\bibfnamefont {N.}~\bibnamefont
  {Pinto-Neto}}, \bibinfo {author} {\bibfnamefont {E.~S.}\ \bibnamefont
  {Santini}}, \ and\ \bibinfo {author} {\bibfnamefont {F.~T.}\ \bibnamefont
  {Falciano}},\ }\href {\doibase 10.1016/j.physleta.2005.06.080} {\bibfield
  {journal} {\bibinfo  {journal} {Phys. Lett.}\ }\textbf {\bibinfo {volume}
  {A344}},\ \bibinfo {pages} {131} (\bibinfo {year} {2005})},\ \Eprint
  {http://arxiv.org/abs/gr-qc/0505109} {arXiv:gr-qc/0505109 [gr-qc]}
  \BibitemShut {NoStop}%
\bibitem [{\citenamefont {Falciano}\ \emph {et~al.}(2013)\citenamefont
  {Falciano}, \citenamefont {Pinto-Neto},\ and\ \citenamefont
  {Vitenti}}]{Falciano:2013uaa}%
  \BibitemOpen
  \bibfield  {author} {\bibinfo {author} {\bibfnamefont {F.~T.}\ \bibnamefont
  {Falciano}}, \bibinfo {author} {\bibfnamefont {N.}~\bibnamefont
  {Pinto-Neto}}, \ and\ \bibinfo {author} {\bibfnamefont {S.~D.~P.}\
  \bibnamefont {Vitenti}},\ }\href {\doibase 10.1103/PhysRevD.87.103514}
  {\bibfield  {journal} {\bibinfo  {journal} {Phys. Rev.}\ }\textbf {\bibinfo
  {volume} {D87}},\ \bibinfo {pages} {103514} (\bibinfo {year} {2013})},\
  \Eprint {http://arxiv.org/abs/1305.4664} {arXiv:1305.4664 [gr-qc]}
  \BibitemShut {NoStop}%
\bibitem [{\citenamefont
  {Brandenberger}(2011{\natexlab{b}})}]{Brandenberger_2011}%
  \BibitemOpen
  \bibfield  {author} {\bibinfo {author} {\bibfnamefont {R.~H.}\ \bibnamefont
  {Brandenberger}},\ }\href {\doibase 10.1088/0264-9381/28/20/204005}
  {\bibfield  {journal} {\bibinfo  {journal} {Classical and Quantum Gravity}\
  }\textbf {\bibinfo {volume} {28}},\ \bibinfo {pages} {204005} (\bibinfo
  {year} {2011}{\natexlab{b}})}\BibitemShut {NoStop}%
\bibitem [{\citenamefont {Gasperini}\ and\ \citenamefont
  {Veneziano}(2003)}]{Gasperini:2002bn}%
  \BibitemOpen
  \bibfield  {author} {\bibinfo {author} {\bibfnamefont {M.}~\bibnamefont
  {Gasperini}}\ and\ \bibinfo {author} {\bibfnamefont {G.}~\bibnamefont
  {Veneziano}},\ }\href {\doibase 10.1016/S0370-1573(02)00389-7} {\bibfield
  {journal} {\bibinfo  {journal} {Phys. Rept.}\ }\textbf {\bibinfo {volume}
  {373}},\ \bibinfo {pages} {1} (\bibinfo {year} {2003})},\ \Eprint
  {http://arxiv.org/abs/hep-th/0207130} {arXiv:hep-th/0207130 [hep-th]}
  \BibitemShut {NoStop}%
\bibitem [{\citenamefont {Ashtekar}\ and\ \citenamefont
  {Singh}(2011)}]{AshtekarSingh2011}%
  \BibitemOpen
  \bibfield  {author} {\bibinfo {author} {\bibfnamefont {A.}~\bibnamefont
  {Ashtekar}}\ and\ \bibinfo {author} {\bibfnamefont {P.}~\bibnamefont
  {Singh}},\ }\href {http://stacks.iop.org/0264-9381/28/i=21/a=213001}
  {\bibfield  {journal} {\bibinfo  {journal} {Classical and Quantum Gravity}\
  }\textbf {\bibinfo {volume} {28}},\ \bibinfo {pages} {213001} (\bibinfo
  {year} {2011})}\BibitemShut {NoStop}%
\bibitem [{\citenamefont {Bojowald}\ and\ \citenamefont
  {Brahma}(2016)}]{BojowaldPRD93}%
  \BibitemOpen
  \bibfield  {author} {\bibinfo {author} {\bibfnamefont {M.}~\bibnamefont
  {Bojowald}}\ and\ \bibinfo {author} {\bibfnamefont {S.}~\bibnamefont
  {Brahma}},\ }\href {\doibase 10.1103/PhysRevD.93.125001} {\bibfield
  {journal} {\bibinfo  {journal} {Phys. Rev. D}\ }\textbf {\bibinfo {volume}
  {93}},\ \bibinfo {pages} {125001} (\bibinfo {year} {2016})}\BibitemShut
  {NoStop}%
\bibitem [{\citenamefont {Ashtekar}\ \emph
  {et~al.}(2006{\natexlab{b}})\citenamefont {Ashtekar}, \citenamefont
  {Pawlowski},\ and\ \citenamefont {Singh}}]{AshtekarPRD73}%
  \BibitemOpen
  \bibfield  {author} {\bibinfo {author} {\bibfnamefont {A.}~\bibnamefont
  {Ashtekar}}, \bibinfo {author} {\bibfnamefont {T.}~\bibnamefont {Pawlowski}},
  \ and\ \bibinfo {author} {\bibfnamefont {P.}~\bibnamefont {Singh}},\ }\href
  {\doibase 10.1103/PhysRevD.73.124038} {\bibfield  {journal} {\bibinfo
  {journal} {Phys. Rev. D}\ }\textbf {\bibinfo {volume} {73}},\ \bibinfo
  {pages} {124038} (\bibinfo {year} {2006}{\natexlab{b}})}\BibitemShut
  {NoStop}%
\bibitem [{\citenamefont {Mielczarek}(2009)}]{MielczarekPLB3}%
  \BibitemOpen
  \bibfield  {author} {\bibinfo {author} {\bibfnamefont {J.}~\bibnamefont
  {Mielczarek}},\ }\href {\doibase
  https://doi.org/10.1016/j.physletb.2009.04.034} {\bibfield  {journal}
  {\bibinfo  {journal} {Physics Letters B}\ }\textbf {\bibinfo {volume}
  {675}},\ \bibinfo {pages} {273 } (\bibinfo {year} {2009})}\BibitemShut
  {NoStop}%
\bibitem [{\citenamefont {Wilson-Ewing}(2016)}]{WilsonEwing2}%
  \BibitemOpen
  \bibfield  {author} {\bibinfo {author} {\bibfnamefont {E.}~\bibnamefont
  {Wilson-Ewing}},\ }\href {\doibase 10.1142/S0218271816420025} {\bibfield
  {journal} {\bibinfo  {journal} {International Journal of Modern Physics D}\
  }\textbf {\bibinfo {volume} {25}},\ \bibinfo {pages} {1642002} (\bibinfo
  {year} {2016})}\BibitemShut {NoStop}%
\bibitem [{\citenamefont {Elizalde}\ \emph {et~al.}(2015)\citenamefont
  {Elizalde}, \citenamefont {Haro},\ and\ \citenamefont
  {Odintsov}}]{Elizalde:2014uba}%
  \BibitemOpen
  \bibfield  {author} {\bibinfo {author} {\bibfnamefont {E.}~\bibnamefont
  {Elizalde}}, \bibinfo {author} {\bibfnamefont {J.}~\bibnamefont {Haro}}, \
  and\ \bibinfo {author} {\bibfnamefont {S.~D.}\ \bibnamefont {Odintsov}},\
  }\href {\doibase 10.1103/PhysRevD.91.063522} {\bibfield  {journal} {\bibinfo
  {journal} {Phys. Rev.}\ }\textbf {\bibinfo {volume} {D91}},\ \bibinfo {pages}
  {063522} (\bibinfo {year} {2015})},\ \Eprint {http://arxiv.org/abs/1411.3475}
  {arXiv:1411.3475 [gr-qc]} \BibitemShut {NoStop}%
\end{thebibliography}%

\end{document}